\documentclass[journal=jctcce,manuscript=article]{achemso}
\usepackage{amsmath}
\usepackage[normalem]{ulem}
\usepackage[table]{xcolor}
\usepackage[numbers,super]{natbib}
\usepackage{soul}

\usepackage{amssymb}
\usepackage{tikz}
\usepackage{xr}
\usepackage{physics}
\usepackage{comment}
\usepackage{graphicx}%
\usepackage{dcolumn}%
\usepackage{bm}%
\usepackage{xcolor}
\newcommand{\SI}{Supporting Information}
\newcommand{\bs}{\ensuremath{\mathbf{s}}}
\newcommand{\br}{\ensuremath{\mathbf{r}}}
\newcommand{\bc}{\ensuremath{\mathbf{c}}}

\newcommand{\bSigma}{\ensuremath{\mathbf{\Sigma}}}
\newcommand{\bmu}{\ensuremath{\boldsymbol{\mu}}}

\newcommand{\bU}{\ensuremath{\mathbf{U}}}

\makeatletter
\newcommand*{\addFileDependency}[1]{%
  \typeout{(#1)}
  \@addtofilelist{#1}
  \IfFileExists{#1}{}{\typeout{No file #1.}}
}
\makeatother

\newcommand*{\myexternaldocument}[1]{%
    \externaldocument{#1}%
    \addFileDependency{#1.tex}%
    \addFileDependency{#1.aux}%
}
\myexternaldocument{SI}

\title{Global free energy landscapes as a smoothly joined collection of local maps}

\author{F. Giberti}
\affiliation{Laboratory of Computational Science and Modeling, Institute of Materials, {\'E}cole Polytechnique F{\'e}d{\'e}rale de Lausanne, 1015 Lausanne, Switzerland}%
\email{federico.giberti@epfl.ch}

\author{G. A. Tribello}
\affiliation{Atomistic Simulation Centre, School of Mathematics and Physics, Queen's University Belfast, Belfast, BT14 7EN, United Kingdom}

\author{M. Ceriotti}%
\affiliation{Laboratory of Computational Science and Modeling, Institute of Materials, {\'E}cole Polytechnique F{\'e}d{\'e}rale de Lausanne, 1015 Lausanne, Switzerland}%
 \email{michele.ceriotti@epfl.ch}
 
 
\date{\today}%

\begin{document}

\begin{abstract}
Enhanced sampling techniques have become an essential tool in computational chemistry and physics, where they are applied to sample activated processes that occur on a time scale that is inaccessible to conventional simulations. 
Despite their popularity, it is well known that they have constraints that hinder their applications to complex problems. The core issue lies in the need to describe the system using a small number of collective variables (CVs). Any slow degree of freedom that is not properly described by the chosen CVs will hinder sampling efficiency. However, exploration of configuration space is also hampered by including variables that are not relevant to describe the activated process under study.

This paper presents the Adaptive Topography of Landscape for Accelerated Sampling (ATLAS), a new biasing method capable of working with many CVs. The root idea of ATLAS is to apply a divide-and-conquer strategy where the high-dimensional CVs space is divided into basins, each of which is described by an automatically-determined, low-dimensional set of variables. A well-tempered metadynamics-like bias is constructed as a function of these local variables. Indicator functions associated with the basins switch on and off the local biases, so that the sampling is performed on a collection of low-dimensional CV spaces, that are smoothly combined to generate an effectively high-dimensional bias. 
The unbiased Boltzmann distribution is recovered through reweigting, making the evaluation of conformational and thermodynamic properties straightforward. The decomposition of the free-energy landscape in local basins can be updated iteratively as the simulation discovers new (meta)stable states. 

\end{abstract}

\maketitle

\section{Introduction}
Enhanced sampling (ES) methods coupled to sampling techniques such as Monte Carlo (MC) and Molecular Dynamics (MD) have become a cornerstone of molecular modeling. While MC and MD are invaluable techniques for investigating the structure and dynamics of matter with an atomistic definition, it is well known that high free-energy barriers hinder their explorative power. In cases where one needs to investigate activated processes, ES methods can be precious as proven by their applications to study phase transitions\cite{francia2020systematic,palm+14nature,khal+11nmat,angi+10prb,piaggi2020phase}, protein folding\cite{piet-laio09jctc,smith2018multi}, chemical reactions\cite{laporte2020formic,fleming2016new,pietrucci2015formamide} and many other phenomena. Many of these ES methods are based on the concept of increasing the sampling of low-probability microstates by introducing a time-dependent external potential\cite{laio-parr02pnas,bard+08prl,pian-laio07jpcb,pfaendtner2015efficient,vals-parr14prl,whitmer2014basis,sidky2018learning,invernizzi2020unified}. The external potential, often called the bias, is expressed as a function of a small number of collective variables (CVs), which are functions of the atomic coordinates. The CVs play many fundamental roles. They decrease the dimensionality of the space needed to be explored and they identify and separate the metastable minima and transition states involved in the activated process one aims to study. The choice of CVs is of paramount importance, but unfortunately, it is hard to decide which and how many should be used.
When selecting the CVs,  there is usually a tug-of-war between using as few CVs as possible and increasing their number to represent the activated process faithfully. Including more CVs can help in describing the mechanism of the process. However, it will also increase the dimensionality of the space that the ES has to explore, making its convergence harder to obtain\cite{bussi2020using} -- even though there are on-going efforts to increase the number of degrees of freedom that can be biased  effectively~\cite{prakash2018biasing,pfaendtner2015efficient}, or to accelerate sampling by selective thermalization of the chosen degrees of freedom rather than by time-dependent biasing~\cite{yu2011temperature,yu2014order}.  

While including additional information certainly helps identify all the different metastable and transition states, it is hard to understand a high-dimensional FES and to use it to construct simple empirical or phenomenological models. Given these problems, there has been a growing effort to create or identify a small number of CVs that could be used to bias an MD or MC calculations. Some notable examples are diffusion maps\cite{rohr+11jcp}, Sketchmap with Field Overlaps Metadynamics\cite{ceri+11pnas,trib+12pnas}, as well as Artificial Neural Networks\cite{sidky2018learning,sidky2020machine}, Deep-LDA \cite{bonati2020data} and RAVE~\cite{ribeiro2018reweighted}. 
These methods aim to achieve a global non-linear dimensionality reduction to find a low-dimensional representation of the CVs. Other methods, such as LDA\cite{mendels2018collective,mendels2018folding} and TiCA \cite{naritomi2011slow,m2017tica}, search for a global dimensionality reduction assuming the existence of a linear discriminant which separates the metastable states. 
Whenever a reasonable guess for a specific reaction pathway is available, an effective dimensionality reduction strategy is to build a one-dimensional reaction coordinate as a path in the high-dimensional CV space~\cite{bran+07jcp,leines2012path}.
A completely different approach has also been attempted, where non-optimized CVs are separated in subsets $\mathbf{K}_X$, and each $\mathbf{K}_X$ is subject to a different biasing potential so that it is easy to converge the independent probability distributions. A few notable examples of these biasing schemes are Bias Exchange \cite{pian-laio07jpcb} as well as Parallel Bias Metadynamics\cite{prakash2018biasing,pfaendtner2015efficient}. 

These methods aim to either sample directly a high-dimensional set of CVs, or to obtain a \emph{global} low dimensional projection, which typically requires complex, difficult to interpret non-linear mappings.  
Rather than aiming to obtain a global low-dimensional description, we introduce a biasing scheme, inspired by metadynamics,\cite{laio-parr02pnas,bard+08prl} whose core idea is to divide the high-dimensional CVs space into local patches and to create a low-dimensional representation of them. In our case, we use a Gaussian Mixture Model (GMM) to partition the high-dimensional space and Principal Component Analysis (PCA) to construct the low-dimensional projections, but other combinations are possible. The total bias is built as a non-linear combination of the local contributions, avoiding the need to create a global low-dimensional manifold. 
The Boltzmann probability distribution can be obtained from the biased simulation using the iterative trajectory reweighting scheme (ITRE) that we recently suggested\cite{gibe+20jctc}, and allows a direct evaluation of free energy differences or FES along any desired degrees of freedom. 
We name this method Adaptive Topography of Landscapes for Accelerated Sampling (ATLAS). 

In what follows, we will first introduce the algorithm, and briefly illustrate how it differs from the state-of-the-art. We will then illustrate how ATLAS is efficient when applied to high-dimensional sampling problems by comparing it with Well-Tempered Metadynamics. To ensure a fair comparison, we apply both methods to a potential with a known analytic form and ensure that the same bias deposition rate is used in both cases. After establishing the effectiveness of ATLAS, we discuss the application to three different atomistic systems, namely a cluster of 38 atoms of Argon, Alanine dipeptide and Alanine tetrapeptide. We take advantage of these cases to discuss an iterative, self-learning strategy to determine an appropriate partitioning of the CV space into local maps, that does not rely on prior information on the system.

\section{Methods}

\subsection{The sampling problem}

MD and MC generate an ensemble of configurations that obey the Boltzmann distribution $P(\br) \approx e^{-\beta U(\br)}$, which depends on the potential energy $U(\br)$ and the inverse temperature $\beta = (k_BT)^{-1}$. Given the probability distribution, any thermodynamic observable can be calculated using
\begin{equation}
\ev{O} = \frac{\int\ d\br \ O(\br)\ e^{-\beta U(\br)}}{\int\ d\br\ e^{-\beta U(\br)}}.
\end{equation}
Unfortunately, the sampling power of these methods is limited. Microstates characterized by a low $P(\br)$ are rarely visited, which is problematic if one wants to investigate a rare (activated) event that involves traversing a region with low $P$. This problem can be ameliorated by introducing a biasing potential $V(\br)$.  
This bias promotes the exploration of regions of phase space that are associated with the process of interest. Since, in general, the optimal bias is not known \emph{a-priori}, most enhanced sampling techniques that rely on a biasing potential build it adaptively, with the aim of discouraging the system from spending too much time in the same region. 
The history-dependent bias potential in these methods is usually expressed as a function of a few selected functions $\bs(\br)$ of the atomic coordinates $\br$. These functions are referred to as collective variables (CVs). Under the action of the bias, the phase-space distribution deviates from the target Boltzmann distribution. In the limit of a slow variation of $V(\bs, t)$, it can be related to the unbiased $P(\br)$ by
\begin{equation}
    \hat{P}(\br,t) = P(\br)\ e^{-\beta [V(\bs(\br),t) - c(t)]}, 
\label{eq:generalP}
\end{equation}
in which 
\begin{equation}
 e^{-\beta c(t)} = \frac{\int\ d \bs\ P(\bs)\ e^{-\beta V(\bs,t)}}{\int d \bs\ P(\bs)},
\label{eq:probab_omega}
\end{equation}
is a time-dependent shift that equalizes the weights of different portions of the trajectory.\cite{bono+09jcc,tiwa-parr14jpcb,valsson2016enhancing}

Once the updating scheme for $V(\bs,t)$ is selected, the sampling efficiency  is mainly related to the choice of CVs $\bs(\br)$. These functions act as a coarse representation of phase space and should be selected to identify and separate the metastable and transition states that characterize the physical/chemical process that one wants want to study. While increasing the number of CVs can improve the description of the rare events we want to sample, it also leads to an exponential increase in the time required to converge the thermodynamic average, because the bias  enhances fluctuations that are not involved in the rare event.
An aspect which is less technical, but not less important, is that even if one could compute a high-dimensional FES, interpreting it would still require discretizing it, and/or projecting it in a lower-dimensional space.  

\subsection{High-dimensional bias with ATLAS}

The philosophy behind ATLAS is to break down the problem of describing a high-dimensional CV space described by a vector $\bs$ of $n_s$ collective variables into $M$ local basins. Within each basin $k$ a lower-dimensional set of CVs $\bc_k(\bs)$ can be defined as a function of the high-dimensional CVs.
The bias acting in each basin is thus low-dimensional, but these local biases are combined to generate an effectively high-dimensional potential, much like a road atlas achieves a complete description of the landscape of a country by breaking it down into small, easy-to-browse pages. 

Similarly to what has been done in reconnaissance metadynamics~\cite{trib+10pnas}, GAMUS~\cite{mara+09jpcb}, OPES~\cite{invernizzi2020rethinking} and GAMBES~\cite{debnath2020gaussian}, we describe basins on the free energy surface in terms of a  Gaussian Mixture Model (GMM):
\begin{equation}
P(\bs) = \pi_0 + \sum_{k=1}^M \pi_k \ G(\bs|\bmu_k,\bSigma_k),
\end{equation}
where each of the $M$ basins is modeled using a normalized Gaussian $G(\bs|\bmu,\bSigma)$, with  mean $\bmu$ and covariance matrix $\bSigma$, and associated with a population $\pi_k$. $\pi_0$ indicates a baseline probability that is meant to encapsulate all states that are not well described by any of the basins. 
We do not, however, use this model directly to build a repulsive bias: most of the time, the actual shape of a free energy basin is not precisely Gaussian, and so a more flexible biasing strategy is needed. 
Instead, we use the GMM to define characteristic functions that identify the regions of the high-dimensional CV space that are to be associated with each basin
\begin{equation}
\theta_k(\bs) =  \frac{\pi_k \ G(\bs|\bmu_k,\bSigma_k)}{\pi_0+\sum_{l=1}^M \pi_l \ G(\bs|\bmu_l,\bSigma_l) },
    \label{eq:pmis}
\end{equation}
This function approaches $1$ when the system resides in a region of CV space associated with the $k$-th basin. A similar expression can be written for the ``background'' basin, 
\begin{equation}
\theta_0(\bs) = \frac{\pi_0}{\pi_0+\sum_{l=1}^M \pi_l \ G(\bs|\bmu_l,\bSigma_l)  }.
    \label{eq:pmi-outside}
\end{equation}
The normalization of these indicator functions ensures that $\sum_{k=0}^M \theta_k=1$, which lends to the $\theta_k(\bs)$ a natural probabilistic interpretation. We refer to them as Probabilistic Motif Identifiers (PMIs) following the nomenclature introduced in Gasparotto \emph{et al.} \cite{gasp-ceri14jcp}, where a GMM was used in a similar spirit to recognize recurring structural patterns in an atomistic simulation. 
The $\pi_k$ parameters for the basins indicate the relative population associated with each cluster and can be obtained from the optimization of an actual GMM model or set to a constant value. The meaning and use of $\pi_0$ is less obvious. One way of rationalizing $\pi_0$ is to imagine that some of the data is not included in any of the $k$ basins.  This data instead belongs to an infinite variance basin with probability $\pi_0$ that we refer to as the ``background basin''. Since the variance of the zeroth basin is infinite, it is not possible to evaluate $\pi_0$ in a simple analytical way.  $\pi_0$ should thus be regarded as an adjustable parameter that determines when the simulation exits the region that is well described by the existing basins and enters ``no man's land''. 

To establish a strategy to select a value for $\pi_0$, one should consider that this parameter has two functions: i) it prevents $\theta_k(\bs)$ from being undetermined when both the numerator and denominator are 0, and ii) it acts as a probability cutoff, so that we do not attribute regions of space where the GMM has a probability smaller than $\pi_0$ to any of the basins. 
Thus, $\pi_0$ should be chosen in such a way that the PMI associated with the background basin, $\theta_0(\bs)$, takes on a sizable value only in regions that are not associated with any of the clusters. A reasonable choice would be to set $\pi_0$ to a fraction $f_0$ (e.g., $95\%$) of the probability assigned to a cluster $k$. This means finding the value of $\pi_0$ for which $\int G(\bs|\bmu,\bSigma) ds = 0.95$.
Recalling that the exponent in a multivariate Gaussian
\begin{equation}
    z = (\bs-\bmu_k)^T\bSigma_k^{-1}(\bs-\bmu_k)
\end{equation}
follows a $\chi^2$ distribution with $n_s$ degrees of freedom, then the value of $\pi_0$ can be readily obtained by evaluating
\begin{equation}
    \pi_0(f_0) = \frac{\pi_k}{\sqrt{(2 \pi)^{n_s} |\bSigma_k|}}\ e^{-z_0^2/2}
    \label{fig:confidenceGMM}
\end{equation}
where $z_0=\operatorname{ICDF_{\chi^2(n_s)}}(1-f_0)$ is the value of the Gaussian exponent that corresponds to the isocontour that discards a fraction $f_0$ of the probability. Since there are $M$ different clusters in our GMM, one can repeat this reasoning for each of the $M$ basins, and select the smallest estimate of $\pi_0$, to avoid disregarding basins that have a low weight.

With these definitions, we can now introduce the ATLAS bias potential
\begin{equation}
V(\bs,t) = \sum_{k=1}^M v_k(\bs,t)\ \theta_k(\bs) + v_0(\bs,t) \theta_0(\bs).
\label{eqn:atlas-bias}
\end{equation}
The first term corresponds to a sum of local biases computed separately in each basin. These are weighted by the indicator functions so that the system feels the bias associated with the $k$-th basin only when it is within the region related to that bias.
The local potential $v_k(\bs,t)$ reads
\begin{equation}
v_k(\bs,t) = h   \sum_{t'=0}^t e^{-V(\bs(t'),t')\Delta T^{-1}} g(\bc_k(\bs)-\bc_k(t')) \frac{ \theta_k(\bs(t'))}{\sum_{l=0}^M \theta_l(\bs(t'))^2}. \label{eq:vk-atlas}
\end{equation}
In this expression, $g(\bc_k-\bc_k(t'))$ is a non-normalized Gaussian function computed for the low-dimensional variables. The indicator functions act so that bias is only added to the basin the system is in at a given time. Note that the denominator in the bias weighting in equation \eqref{eq:vk-atlas} contains the square of the PMIs. Even though the PMIs are themselves normalized in a $L^1$ sense, the contributions to $v_k(\bs,t)$ in equation \eqref{eqn:atlas-bias} are multiplied by $\theta_k(\bs)$. A further $L^2$ normalization is needed to ensure that the work done on the system by the time-dependent bias is independent of the basin decomposition. In other words, every time a set of Gaussian hills are added, the bias increases by a factor $h\ e^{-V(\bs(t'),t')\Delta T^{-1}}$, as it would in a conventional Well-Tempered Metadynamics simulation.

The term $v_0$ in equation \ref{eqn:atlas-bias} corresponds to an adaptive wall that pushes the system back into the region described by the GMM if it spends too much time outside of it. It has a formulation that is identical to~\eqref{eq:vk-atlas}, but as it refers to the ``background basin'' which has infinite variance, we define no CVs for it and we set $g(\bc_0(\bs))\equiv 1$ so: 
\begin{equation}
v_0(\bs,t) =\ h   \sum_{t'=0}^T e^{-V(\bs(t'),t')\Delta T^{-1}} \frac{ \theta_0(\bs(t'))}{\sum_{l=0}^M \theta_l(\bs(t'))^2}.
\end{equation}
Due to the well-tempered damping factor, this term increases more slowly as the simulation proceeds. The convergence of the bias at the boundary of the GMM is thus ensured, and there is a smooth bias over the entirety of CV space. Note, however, that the $v_0(\bs,t)$ bias is constant since it does not depends on $\bc$, and so it can provide a significant force only when the system lies in the transition region at the edge of the GMM.

\begin{figure}
    \centering
    \includegraphics[width=0.5\textwidth]{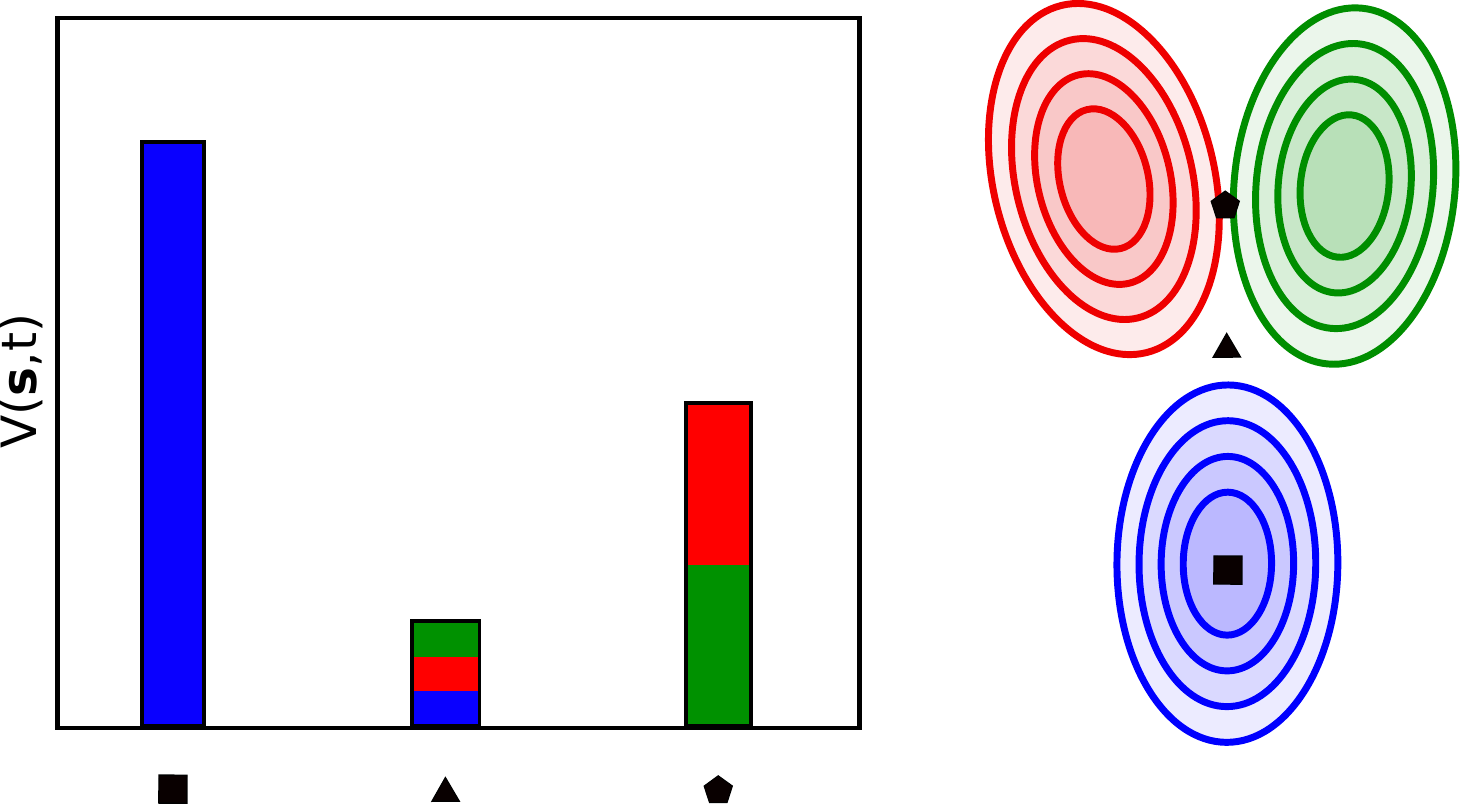}
    \caption{Illustration of how the bias is constructed in ATLAS for three different minima. Although this is just a sketch, it conveys how the total bias is constructed starting from the local bias. The three points, illustrated with three different polygons, are subject to the total potential acting on them, which is the sum of the three local potentials of the red, green, and blue minima. The amount with which each minimum contributes to the potential is sketched in the histogram on the left. For this case, we assumed that no wall was applied to the potential.}
    \label{fig:atlas_sketch}
\end{figure}

In practice, this formulation of the potential switches the local contribution $v_k(\bs,t)$ to the bias on if configuration $\bs(t)$ and $\bs(t')$ are in the same minimum. If the PMIs have a fractional value, then the global potential is obtained as a combination of the local contributions. The form of the ATLAS bias is illustrated in figure \ref{fig:atlas_sketch}, which indicates the bias composition at three different points, with contributions originating from a GMM composed of three basins.

The local bias $v_k(\bs,t)$, is constructed as a combination of Gaussians employing the $\bc$ variables, which are a local projection of the $\bs$ CVs. The $\bc$ functions can be constructed in many different ways. One possibility, which we borrow from Ref.~\citenum{trib+10pnas}, uses the Mahalanobis distance
\begin{equation}
    d_k = \sqrt{(\bs-\bmu_k)^T\bSigma_k^{-1}(\bs-\bmu_k)}
\end{equation}
as a one-dimensional CV.
Another possible choice is the projections of $\bs$ over the $l$ largest eigenvectors of the $\bSigma_k$ matrix, normalized by their eigenvalues $\lambda_l$
\begin{equation}
     c_k^l = \frac{\bs^{T} \bU_k^l}{\sqrt{\lambda_k^l}},
 \end{equation}
with $\bU_k$ being the unitary matrix that satisfies $\bU_k^T\bSigma_k\bU_k=\mathbf{\Lambda}_k$. 
The two methods can also be combined to construct a bi-dimensional space where the first CV is the projection on the largest principal component of $\bSigma_k$, and the second CV is the Mahalanobis distance calculated using the $n_s-1$ remaining components of $\bSigma_k$. In this case the effective dimensionality of the space where the $g(\bc_k(\bs)-\bc_k(t'))$ Gaussians are added is determined by the number of $c_k^l$ used. The optimal number of low-dimensional CVs could be found by looking at the eigenvalue decomposition of $\bSigma_k$, and selecting the eigenvectors with the largest weights. This leave us the freedom to decide how much information to incorporate in the low-dimensional space as well as its size.

Notice how, even though the number of high-dimensional variables can be in principle increased at will -- and one can use some of the many types of structural descriptors commonly used in enhanced sampling calculations, such as angles, angular parameters, inter-atomic distances -- it is clear that the quality of the $\bc_k$ collective variables depend on the initial choice of the $\bs$ features. For example, it is better to avoid introducing redundant information, i.e. CVs that do not contribute in the identification of minima, and can only add noise to the method, as well as CVs that  are inconsistent with fundamental invariances of the system, such as the Cartesian coordinates for which the same minima can be obtained by rotations, or a permutation of atom indices.

Among the many possible approaches one could use to define the local coordinates $\bc_k(\bs))$, we focus on those using the principal components. We refer to methods that only use the top $N$ principal components as $N$D-PCA.  RES indicates that the  distance computed from the residual components of $\bSigma_k$ is used as well as the the principal components when defining the $c$ variables.
The ATLAS formulation of the bias has several desirable features. The first immediate one is that the potential acts in a high dimensional space $\bs$, but is defined and accumulated in small dimensional spaces $\bc_k$. The method does not scale exponentially in the number of CVs, but is instead linear in the number of Gaussians basins $M$ once the $\bc_k$ are defined. Furthermore, the baseline potential $v_0$ provides an adaptive, high-dimensional wall that restraints the system in the relevant region of phase space, and can be used to detect when a new basin has been discovered that does not correspond to any of the $M$ states included in the GMM.

As a closing remark, we would like to discuss the differences between ATLAS and methods such as OPES and GAMBES as well as GAMUS\cite{invernizzi2020rethinking,mara+09jpcb,debnath2020gaussian}. These methods use a Gaussian Mixture Model (GMM) to estimate P($\bs$), and then create a static biasing potential, following a schema similar to umbrella sampling. The GMM is then periodically re-estimated, and the bias adjusted multiple times until convergence of the FES or P($\bs$) is obtained. If the basins themselves don't have a Gaussian shape, multiple Gaussians need to be added (and multiple iterations of the scheme need to be performed) before a flat bias is achieved, and so the scaling with dimensionality is comparable to that of conventional metadynamics. On the other hand, ATLAS, only uses the GMM to partition phase space and define a local coordinate system. The form of the bias within each local topographic description is entirely non-parametric.

\subsection{Post processing}
At the end of the calculation, the unbiased probability distribution P($\bs$) can be obtained by reweighting. While $c(t)$ can be calculated in many ways, we believe that ITRE is the best choice to estimate it in this context, as it does not depend on the number of CVs used. Once $c(t)$ has been obtained, it is possible to evaluate the free energy as a function of any CVs, even those not included in the sampling. 

It is also straightforward to calculate free energy differences between two regions $a$ and $b$ given the GMM. The probability of being in one region can be easily obtained from the PMIs by using
\begin{equation}
    P_k = \frac{\int_0^T \theta_k(\bs(\br(t))) e^{\beta(V(\bs(t),t)-c(t))} dt}{\int_0^T e^{\beta(V(\bs(t),t)-c(t))} dt}
    \label{eq:P_of_k}
\end{equation}
This expression does not depend on $\bs$ and does not require any parameter to identify the region $k$. The free energy difference between two basins can be obtained as usual from:
\begin{equation}
    \Delta G_{ab} = -kT \log\frac{P_a}{P_b}
    \label{eq:delta_g}
\end{equation}
We want to remark that while it is appealing for the function $\theta_k$ to represent a basin, i.e., a stable structure of the FES, this is by no means a necessity. The GMM can identify as a cluster an unstable or a region where there is  lot of degeneracy in the value of the FES, such as a large-entropy basin. This does not decrease the efficacy of equation \eqref{eq:delta_g}, which measures the difference in population between two regions and can thus be used to evaluate the FES's convergence.

An implementation of ITRE capable of reweighting ATLAS calculations can be found in the \textit{hack-the-tree} branch of PLUMED-2.0\cite{trib+14cpc}\footnote{The most recent ATLAS implementation can be obtained at the commit with SHA-1 92086a691252ac862e52de659a37ad88cce68c5c.}, as well as in a python module in the cosmo-tool repository of the COSMO research group github available at \url{https://github.com/cosmo-epfl/cosmo-tools}%{https://github.com/cosmo-epfl/cosmo-tools}.
%MC TODO before resubmission ?????
%The input script to regenerate the calculations illustrated in the manuscript have been deposited on the PLUMED-NEST repository and are identified with the code XXXXXXX \cite{bonomi2019promoting}.

\subsection{Iterative determination of the local maps}

\begin{figure}
    \centering
    \includegraphics[width=0.5\textwidth]{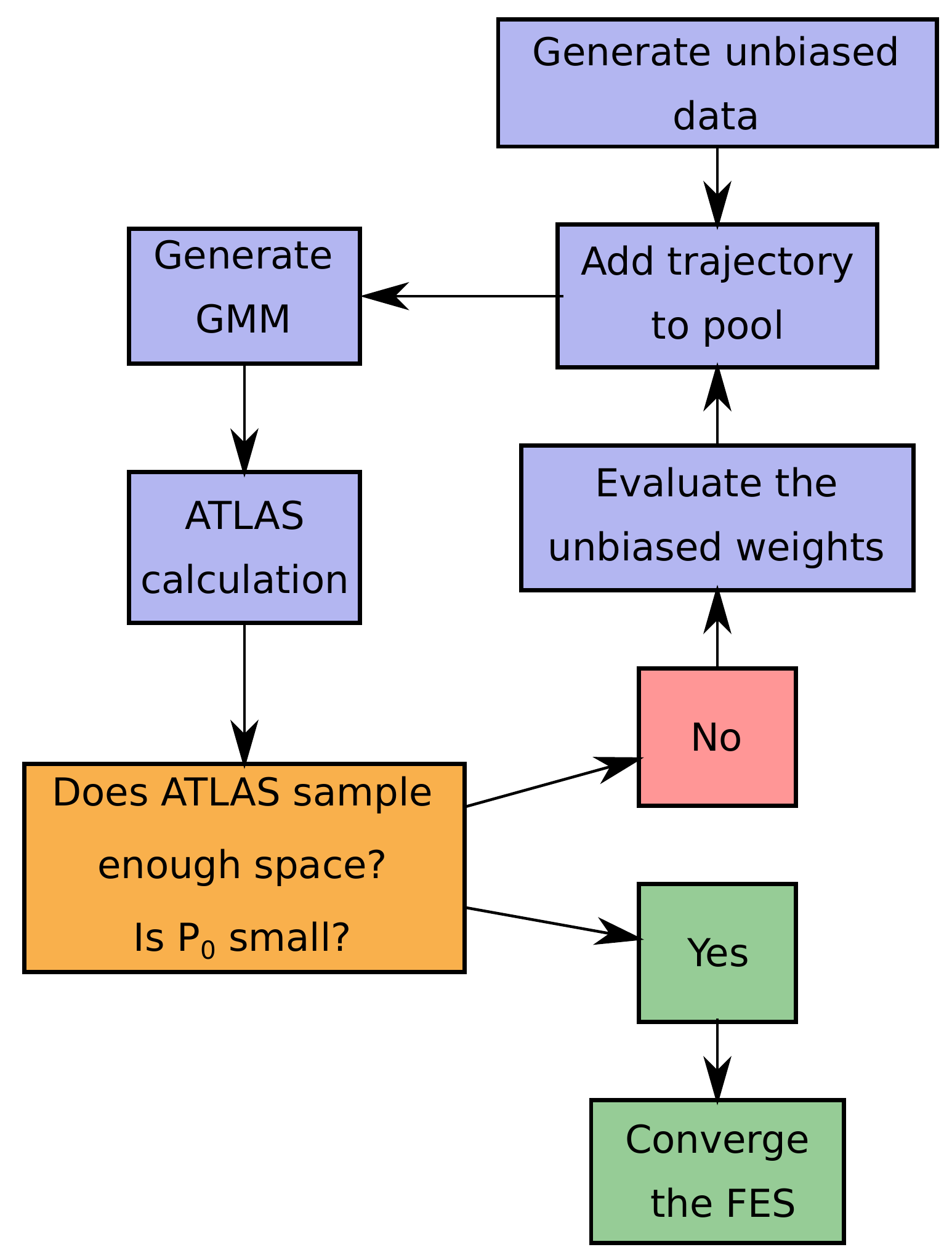}
    \caption{Diagram illustrating an iterative scheme to update the GMM that underlies ATLAS.}
    \label{fig:learning-protocol}
\end{figure}

Even though in Section~\ref{sec:model-pots} we will assume that the GMM that underlies the space partitioning in ATLAS is known, it is relatively simple to extend the method to include a self-learning biasing scheme,  which is summarized in Figure~\ref{fig:learning-protocol}.
One starts by building a pool of biased or unbiased trajectories, that do not need to sample the entire phase space. Each trajectory is associated with weights (computed with ITRE for trajectories with a time-dependent bias), combined with the others, and used to fit a GMM. 
Based on this GMM, an ATLAS simulation is run, and the exploration of CV space is monitored so a decision can be made about when to update the GMM.  We find that a rather effective way to detect that the ATLAS GMM needs updating is to monitor the population $P_0$ associated with the ``background basin'', computed according to Eq.~\eqref{eq:P_of_k}. If $P_0$ becomes substantial, it indicates that the system has found a new (meta) stable state, that is not described by any of the GMM basins.
When this condition is satisfied, the simulation is stopped, the pool of trajectories is updated by adding the new ATLAS-biased trajectory, and the entire pool is used to train a new GMM.
Once sampling can continue without observing a persistent increase of $P_0$, or without other signs of hysteresis or sampling inefficiency, the ATLAS simulation can be continued until the FES is converged.

Pooling the trajectories from different iterations ensures that the method is robust, and each iteration does not need to be converged thoroughly.  The GMM also does not lose the memory of states that have been visited in only some of the calculations. 
However, particularly during early iterations where ATLAS is based on highly incomplete sampling, it is advisable to restart sampling without keeping the previous bias active, to avoid introducing artifacts in the bias, that would complicate convergence of later runs. 
Note also that the stopping criterion for the iteration is based on general considerations, and in practical cases one might use more specific requirements to ensure  convergence of the GMM and the FES.

\section{Model potentials}\label{sec:model-pots}

\begin{figure}
    \centering
    \includegraphics[width=\textwidth]{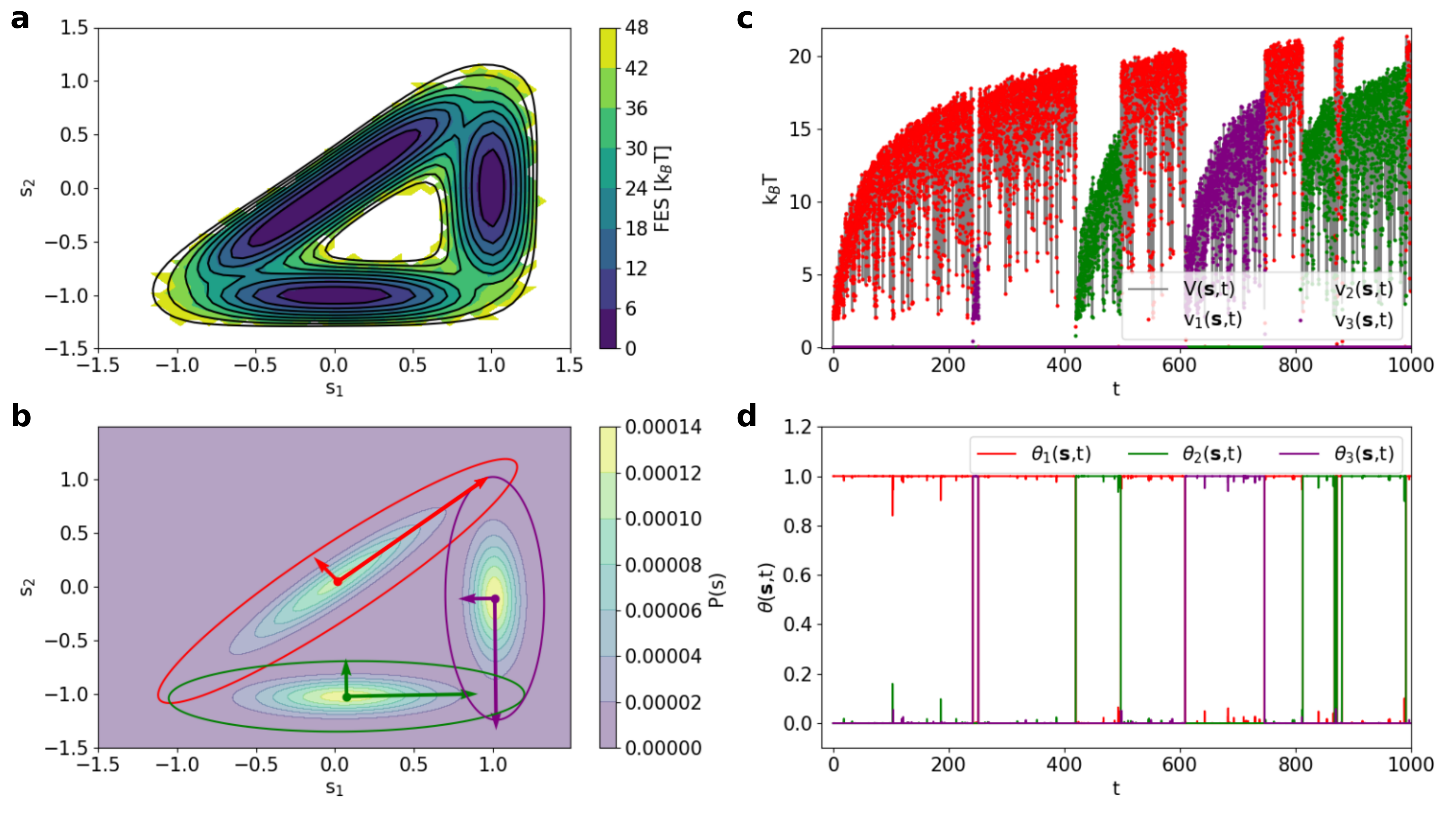}
    \caption{Illustration of the ATLAS framework. (A) The FES acting on the Langevin particle. The solid black contours correspond to the FES estimated using parallel tempering and matches the estimate obtained by reweighting an ATLAS trajectory, represented with filled contours, color-coded according to the indicated scale perfectly. (B) The GMM used is fitted to the probability distribution for a Langevin particle moving on the FES. For each cluster, we show the two principal components, as well as a contour enclosing 99$\%$ of the probability for each Gaussian. (C) Total bias potential $V(\bs,t)$ along an ATLAS trajectory, together with the contributions from individual basins $v_k(\bs,t)$, color-coded according to the  scheme in panel B. (D) PMIs for the three basins along the same trajectory as in panel C.}
    \label{fig:example_pot}
\end{figure}

Before analyzing ATLAS performance in detail, we want to present a practical example that will help the reader better understand the algorithm and its parameters. 
We estimate the FES for a single particle following a Langevin dynamics on the 2D potential illustrated in panel A of figure \ref{fig:example_pot}. 
The details of the MD for this simulation are summarized in the \SI. The potential consists of three very narrow and anisotropic minima, separated by barriers of roughly $\approx$20 k$_B$T, and its probability distribution can be described using a GMM with M=3, as illustrated in panel B of figure \ref{fig:example_pot}. Each minimum is identified with a different color, and the two principal components are also indicated as arrows starting from $\bmu_k$. 
The colored ellipses are drawn using equation \eqref{fig:confidenceGMM} so that they enclose 99$\%$ of the probability of each Gaussian cluster. As can be seen in panels C and D, the simulation starts in the third minimum. After $\approx$ 400 Gaussian depositions $\tau$, the trajectory escapes the initial basin and visits a second minimum. The sequence with which the minima are visited is, of course, irrelevant. However, we want to draw the reader's attention to the fact that when the system changes minimum, the contributions from the three local potentials switch on or off following the value of the PMIs reported in panel D of figure \ref{fig:example_pot}. The local potentials match the total potential $V(\bs,t)$ almost precisely, because in this simple example the PMIs switch sharply between basins. Small differences, notable as a few isolated dots in panel C, arise when the system jumps from one minimum to another.  At those points the system is in a region where two GMM clusters overlap, and the PMIs take a fractional value.
The FES obtained at the end of the calculation is illustrated in panel A as filled contours, while the reference, obtained with Parallel Tempering, is drawn as solid black lines. A movie illustrating the first steps of this calculation can be found in the \SI.

\subsection{A challenging free energy surface}

Having shown a practical example of how ATLAS constructs the potential, we can now investigate the method's performance when increasing the dimensionality $D$. 
We introduce a construction of a $D$-dimensional potential with $D+1$ minima, which generalizes that used in Ref.~\citenum{gibe+20jctc}, designed so that the basins do not overlap with each other, and so that each transition involves a different combination of the $D$ spatial dimensions. The basins are arranged in a loop so that it is possible to traverse them sequentially by going across high free-energy barriers and returning to the starting point. 
The analytical form of the potential is quite complex and is described in detail in the \SI. This family of FES is designed to be intelligible, but to challenge accelerated sampling algorithms in every possible way, and we refer to it as the $D$-dimensional FES from hell, $D$-FFH. 

\begin{figure}
    \centering
    \includegraphics[width=\textwidth]{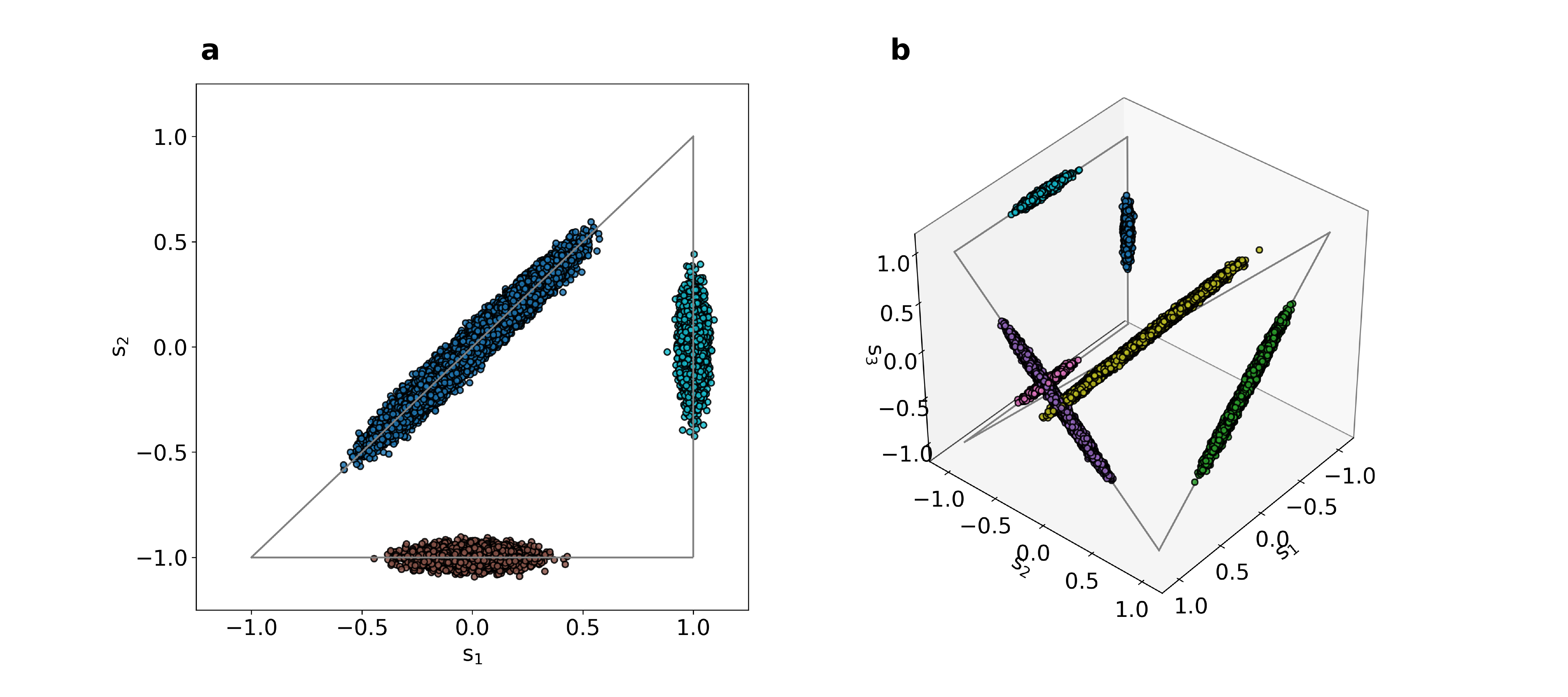}
    \caption{Examples of the circular topology used in the single Langevin particle systems for the 2D (panel A) and the 3D cases (Panel B). The points belonging to the different minima are colored differently, and a gray line helps the reader to visualize the circular topology.}
    \label{fig:topology_examples}
\end{figure}

We consider examples with $D=2,3,6$.
The $D=2$ FFH corresponds to the toy example discussed in the previous Section. For $D=3$, we complicate the general construction further to include six basins arranged in a circular topology (Fig.~\ref{fig:topology_examples}). Finally, for $D=6$ we use the general construction, that corresponds to $D+1=7$ minima. 
We directly compare ATLAS and a well-established method that the majority of the community is familiar with, i.e., Well-Tempered Metadynamics \cite{bard+08prl}. 

The GMMs needed to perform the ATLAS calculations, and the references used to evaluate the convergence of the FES, were obtained by running unbiased parallel tempering calculations with i-PI as server and PLUMED-2.0 as a driver\cite{ipicode,kapi+19cpc}. 
We tested three different flavors of the ATLAS scheme. The first uses the first principal component of the covariance matrices (1D-PCA), the second uses the two principal components (2D-PCA), and the last employs the first principal components as well as the distance from the associated 1D subspace (RES). The concept behind RES is to push the system in the direction along which the minimum is oriented and at the same time far from it, similar to the spirit of path collective variables~\cite{bran+07jcp}. 
The bias is accumulated on a 1D (1D-PCA) or 2D grid (2D-PCA and RES), whose limits are chosen to span the region in $\bc_k$ for which the PMIs are non-negligible. 
The Well-Tempered Metadynamics calculation (META) performed for the 2D, and 3D systems are also performed using a grid. Unfortunately, applying a grid in the  6D case would be impossible, and so for this case only, we did not employ a grid and computed the metadynamics bias by explicitly summing over the history of the bias. For both the Metadynamics and ATLAS calculations, we use the same hills height and deposition rate so that the work performed on the system can be readily compared. We perform a total of 12 independent META, 1D-PCA, 2D-PCA, and RES trajectories for each system, to be able to estimate the uncertainty of the resulting FES as the standard error in the mean.

At the end of the calculations, the unbiased $P(\bs)$ and the associated FES are reconstructed using ITRE with a stride of 10 depositions. Fewer than five iterations used are needed to converge $c(t)$. We calculate all the $D$ mono-dimensional and all the $D(D-1)/2$ bi-dimensional P($\bs$) from the weighted microstates to enumerate all the possible mono and bi-dimensional FES combinations. To estimate the efficiency of the different methods, we use two different metrics, the Kullback-Leibler divergence $D_{KL}$ and the free energy differences between basins $\Delta F_{ab}$. The former is defined as
\begin{equation}
    D_{KL} = \int \tilde{P}(\bs) \log \left ( \frac{\tilde{P}(\bs)}{P(\bs)} \right ) d \bs,
    \label{eq:KL_div}
\end{equation}
where $\tilde{P}(\bs)$ is the reference probability evaluated by parallel tempering. This can be recast as  
\begin{equation}
    D_{KL} = \beta \int \tilde{P}(\bs)\ \left ( F(\bs) - \tilde{F}(\bs) \right ) d \bs,
\end{equation}
which illustrates how the KL divergence measures the difference between the reference FES and the reweighed FES, weighted by the reference probability $\tilde{P}(\bs)$. The second metric used to compare the methods is the Free Energy Difference $\Delta F_{ab}$ between two different minima $a$ and $b$ evaluated following equations \eqref{eq:P_of_k} and \eqref{eq:delta_g}.
While $D_{KL}$ is a global metric, as it is evaluated as an integral over the $\bs$ space, $\Delta F_{ab}$ is a more ``local'' metric since it only depends on the accuracy of sampling of the relative populations of $a$ and $b$. 

\begin{figure}
    \centering
    \includegraphics[width=\textwidth]{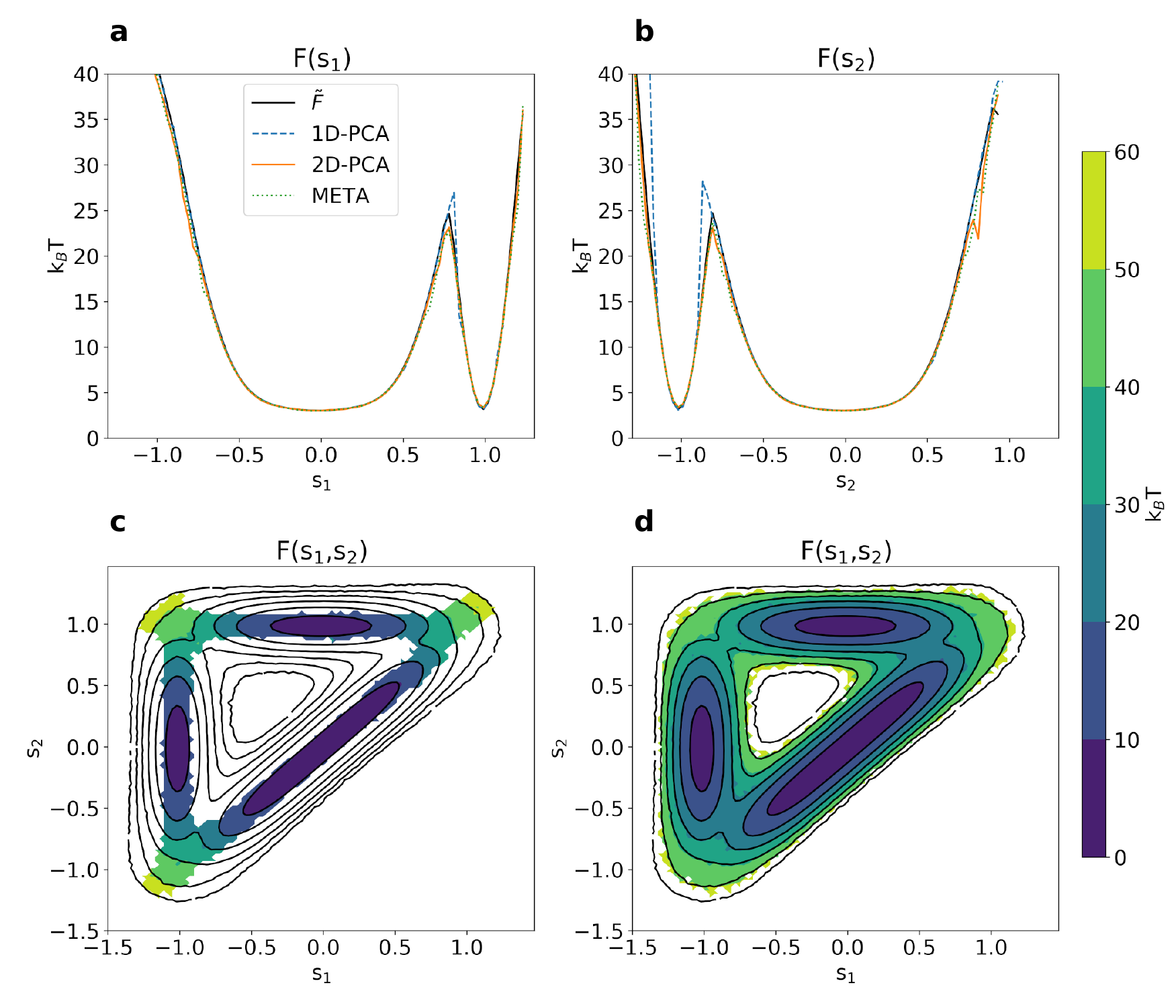}
    \caption{Results obtained with the ATLAS and Metadynamics methods, compared to the parallel tempering references, for the 2D FFH. In Panel A) and B), we reported the reference FES, ATLAS, and META results, as a function of $s_1$ and $s_2$. The 2D FES for 1D-PCA and 2D-PCA have been reported in panel C) and D), respectively. While META and 2D-PCA provide a similar quantitative result, 1D-PCA cannot accurately sample the barriers between minima. Still, it is capable of quantitatively reproducing the shape of the minima, both in 1D and in 2D.}
    \label{fig:2D_system_FES}
\end{figure}

\begin{figure}
    \centering
    \includegraphics[width=\textwidth]{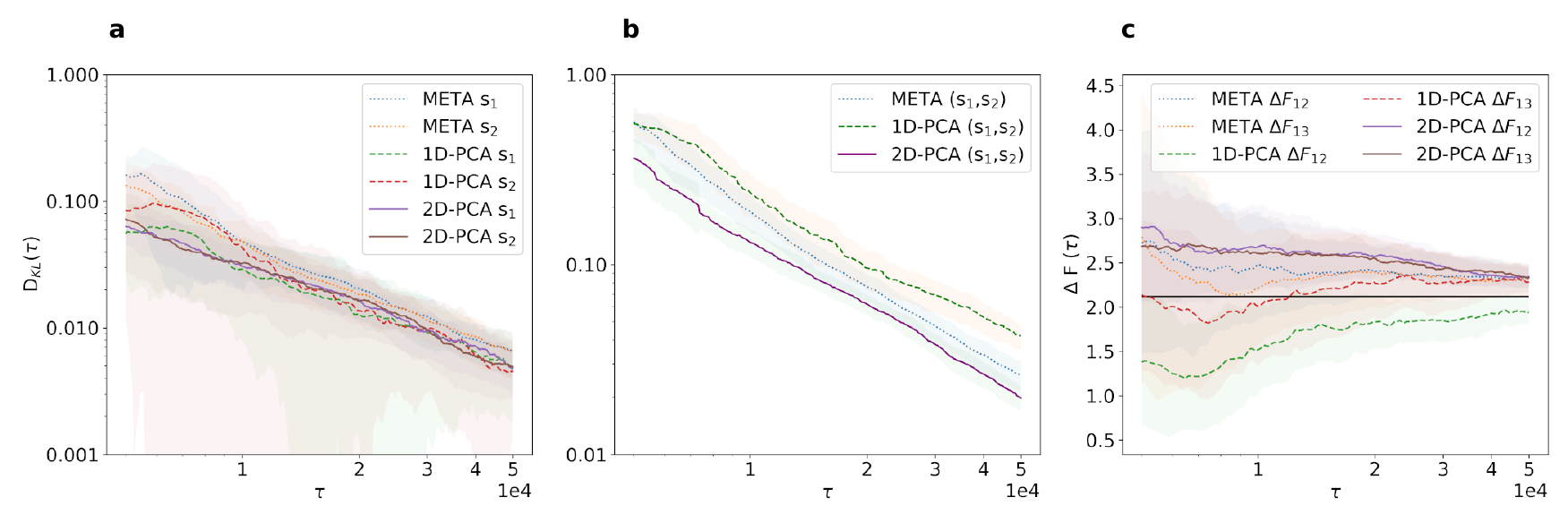}
    \caption{Estimate of the convergence of the FES as a function of time for the three ES methods. (A) Convergence of $D_{KL}$ for 1D projections of the FES. (B) Convergence of $D_{KL}$ for the 2D FES. (C)  Convergence of $\Delta F_{ab}$ between pairs of basins, using the basin centered in $(0,0)$ as reference. }
    \label{fig:2D_system_comparison}
\end{figure}

For the $D=2$ system, we compare 1D-PCA,  2D-PCA, and META calculations, using the results from PT as the ground truth. The results obtained for this system are illustrated in figure \ref{fig:2D_system_FES}. 
META and 2D-PCA provide very similar results (so similar that we report the META results only in the SI).  This similarity is unsurprising given that they both sample the full-dimensional space, although with a different metric. 1D-PCA, on the other hand, samples only the bottom of each basin in the narrow direction, which is by design: only the first principal component is biased. The transition state behavior's deserves further discussion: the 1D coordinate is not sufficient to describe displacements in the region between two basins, and so most of the trajectories ``overshoot'' and follow a higher-energy transition path. 
It is remarkable that despite this limitation, ATLAS recovers very accurately the shape and relative free energy of the various minima, even though there is a rather high error in the transition state region.
The timescales with which the three methods reconstruct the different FESs are also quite similar, as can be seen in figure \ref{fig:2D_system_comparison}. In fact, 1D-PCA performs only slightly worse than the full-dimensional sampling methods, which indicates that even though there are inaccuracies in the reconstruction of the transition states, there is no substantial hysteresis.

\begin{figure}
    \centering
    \includegraphics[width=0.85\textwidth]{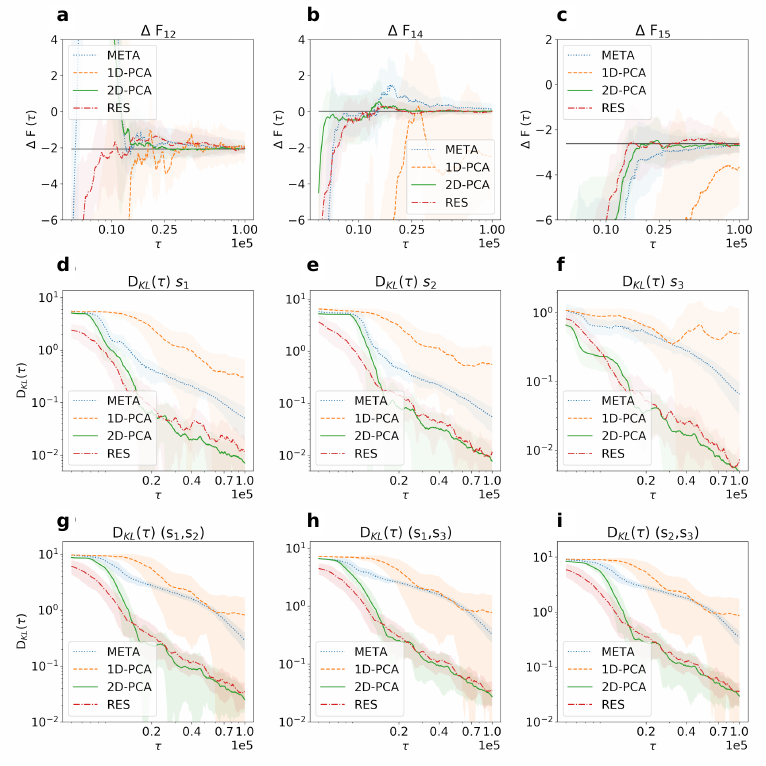}
    \caption{Comparison between the Metadynamics 1D-PCA, 2D-PCA, and RES for the 3D system. The free energy differences $\Delta F$ between the first and second, fourth, and fifth minima are illustrated in panel (A), (B) and (C), respectively. All the methods but 1D-PCA converge to the Parallel Tempering references reported as solid black lines. There is little to no difference in how fast 2D-PCA, RES, and META converge as a function of the bias deposited, with 2D-PCA and RES having a lower variance than META.  The Kullblack-Leibler Divergence D$_{KL}$ for the 1D P($\bs$) are illustrated in panel (D), (E), and (F) and for the 2D P($\bs$) in panel (G), (H) and (I) respectively. As expected, the 1D-PCA shows a poorer convergence compared to other methods. However, 2D-PCA and RES converge faster than META. The FES for the 3D system as well as the basins are reported in figure \ref{fig:3D_box}.}
    \label{fig:3D_fes}
\end{figure}

\subsection{Three dimensional free energy surface}

For the 3D extended FFH we obtain results that are broadly compatible with the observations in 2D. However, the benefits of using a reduced dimensionality in the sampling within the basins become more evident. 
As shown in Figure~\ref{fig:3D_fes}, all the methods yield converged FES in the minima, but the 1D-PCA and (to a lesser extent) RES methods show sub-optimal convergence in the TS region. A more quantitative analysis of the convergence speed (Fig.~\ref{fig:3D_fes}) demonstrates that, judging on the convergence of both $D_{KL}$ and $\Delta F_{ab}$, the 1D version of ATLAS is slowed down by the inadequate description of the transition states, but both the 2D-PCA and RES flavors of ATLAS outperform 3D metadynamics by far. These methods achieve errors that are an order of magnitude smaller than META, for the same simulation time. 
Further convergence tests (reported in the SI) are consistent with this picture. A too aggressive dimensionality reduction hampers the efficiency of phase space exploration, but a more balanced 2D-PCA scheme achieves a very substantial improvement in sampling efficiency.

\begin{figure}
    \centering
    \includegraphics[width=0.85\textwidth]{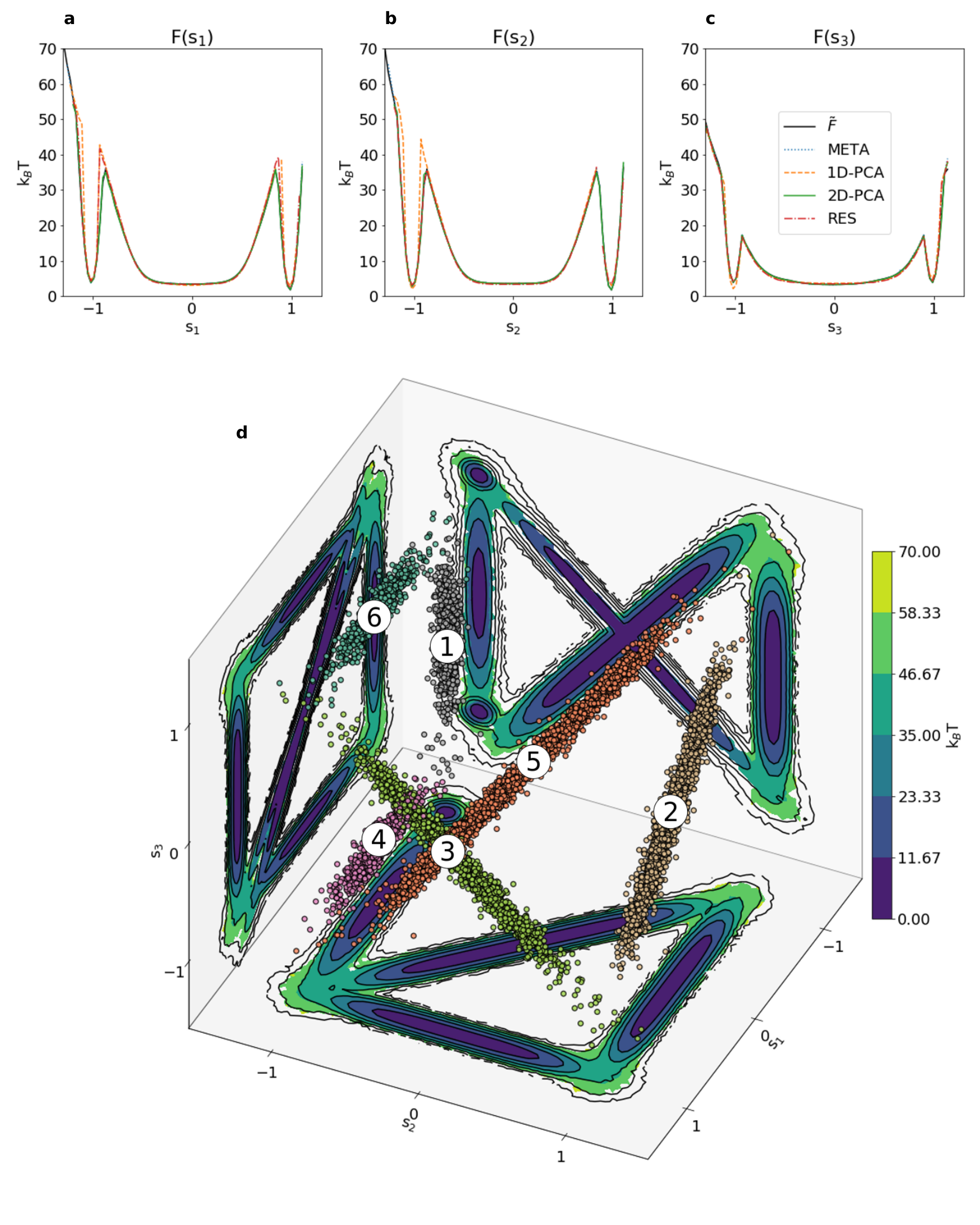}
    \caption{Panel (A), (B) and (C) illustrate the FES for $s_1$, $s_2$ and $s_3$ respectively. As for the 2D cases, META and 2D-PCA faithfully represent the 1D FES, while 1D-PCA and RES cannot represent the barrier for the FES of $s_1$ and $s_2$ properly. Panel (D) illustrates the distribution of the 6 minima in the 3D system, as well as the three 2D FES F($s_1$,$s_2$), F($s_1$,$s_3$), and F($s_2$,$s_3$) obtained with the 2D-PCA method. The free energies for the 2D FESs obtained with Metadynamics are illustrated in the \SI, as well as those obtained with 1D-PCA.}
    \label{fig:3D_box}
\end{figure}

\subsection{Six dimensional free energy surface}
\begin{figure*}
    \centering
    \includegraphics[width=\textwidth]{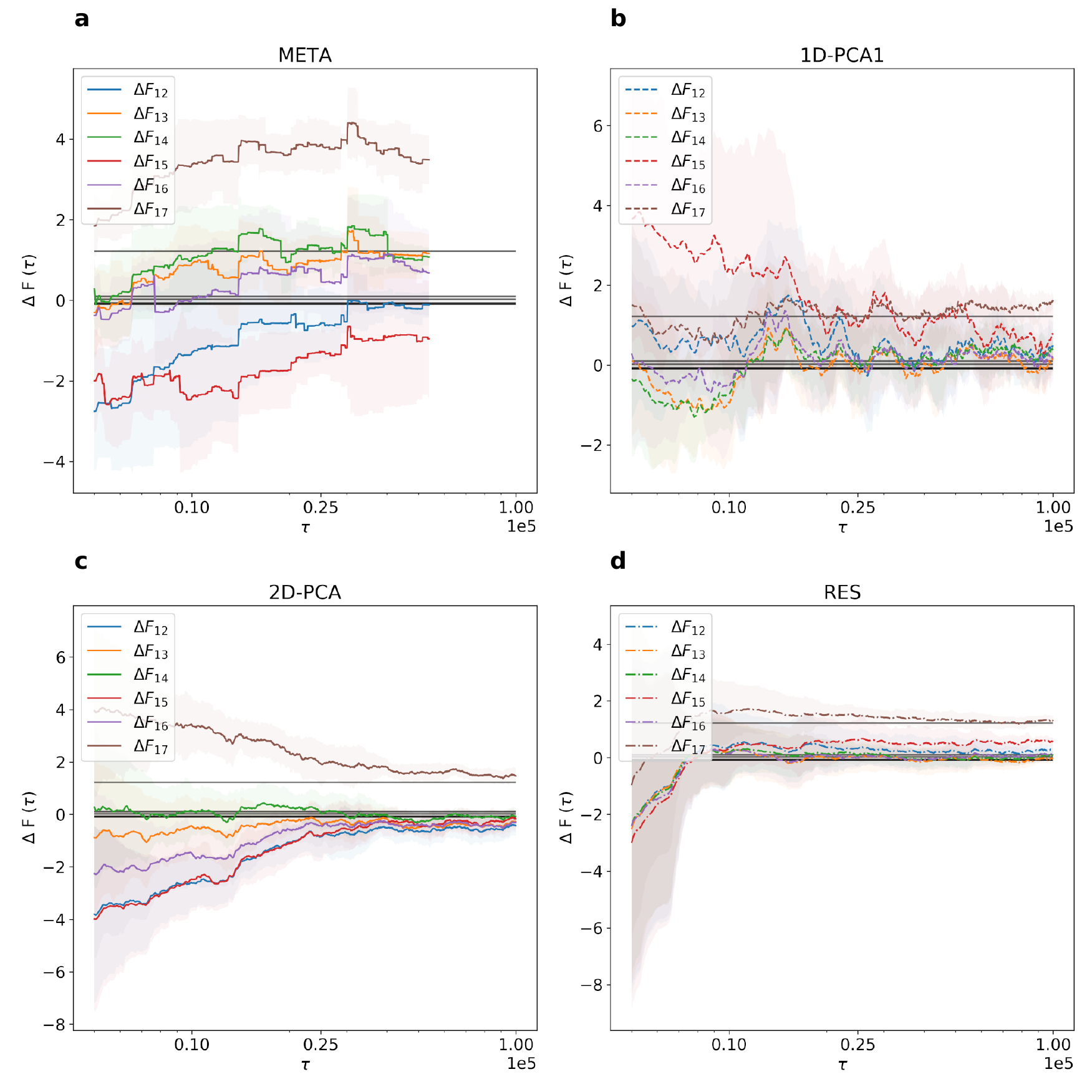}
    \caption{The $\Delta F_{ab}(\tau)$ for the 6D FFH. META, 1D-PCA, 2D-PCA and RES are illustrated in panel A), B), C) and D) respectively.}
    \label{fig:6D_comparison}
\end{figure*}

The difference between the rate at which ATLAS and META sample phase space becomes even more pronounced as the system's dimensionality increases. For the 6D FFH (Figure \ref{fig:6D_comparison}), which contains seven minima arranged along a complicated cyclic path, META calculations struggle to converge the free energy difference between the seven minima. All the flavors of ATLAS reproduce qualitatively the results obtained from the PT calculations, although one can see clearly that 1D-PCA shows a very noisy behavior, and that $\Delta F_{15}$ seems to converge to a slightly incorrect limit with RES local variables. 
ATLAS's convergence trends are comparable to the 3D example, indicating that the sampling behavior depends on the number of minima identified in the GMM rather than the number of high-dimensional CVs.

\begin{figure}
    \centering
    \includegraphics[width=\textwidth]{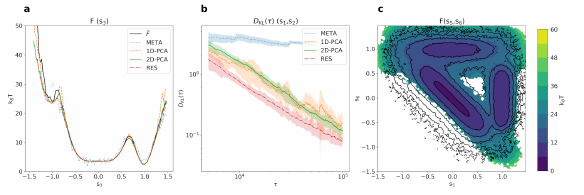}
    \caption{A summary of the diagnostic results for the 6D FFH, comparing sampling with METAD, ATLAS using 1D-PCA, 2D-PCA and RES, and a parallel tempering reference. (A) 1D projection of the FES along the direction $s_3$.  
    (B) Convergence of $D_{KL}(\tau)$ for the 2D FES $F(s_1,s_2)$. (C) An example of a 2D projection of the FES $F(s_5,s_6)$ recovered by reweighting a 2D-PCA ATLAS trajectory. The ATLAS FES is illustrated with colored contours while the PT reference is indicated with black contours.}
    \label{fig:6D_example}
\end{figure}

A thorough analysis of the convergence of the different methods is shown in the SI. Figure~\ref{fig:6D_example} provides a representative example, demonstrating the convergence of 1D and 2D FES projections, as well as a 2D FES obtained with the 2D-PCA framework. 
It is clear from the figure that the three flavors of ATLAS sampling can recover the FES with an error which is much smaller than $k_BT$, even for the high free energy minimum at $\approx25 k_BT$. META can only recover the low free energy states, with a very high level of noise. The filling rate of the basins is also lower, which hinders sampling of the highest free-energy basins. The difficulties of META can also be seen from the extremely slow convergence of the KL divergence, shown in panel (B) for the FES along $s_1,s_2$. We illustrate the 2D FES $F(s_5,s_6)$ obtained with 2D-PCA in panel (C), showing that ATLAS accurately samples the free energy minima and the transition state regions, while -- depending on the projection -- it does not explore the fluctuations in the uninteresting directions. 
In interpreting this comparison, it is important to keep in mind that running a 6D Metadynamics calculation using a grid would be impossible due to memory requirements. Since the time required to run Metadynamics without a grid scales as the square of the trajectory length, the actual computational cost of a META calculation is substantial. Therefore, we had to use only  4 META trajectories, which contributes to the larger error bars in Fig.~\ref{fig:6D_comparison}. It is clear that the improvement in sampling performance enabled by ATLAS is dramatic, even neglecting the computational effort due to the $T^2$ scaling.

\section{Atomistic systems}

The FFH models are challenging test cases for any enhanced sampling algorithm and were specifically designed to exacerbate the difficulties associated with sampling a high-dimensional free-energy landscape. 
To assess how ATLAS performs in real-life applications, we consider three systems that, although simple, are representative of the challenges that are often found in atomistic simulations: i) a cluster of 38 atoms interacting through the Lennard-Jones potential (LJ-38)~\cite{wale03book}, ii) Alanine dipeptide and iii) Alanine tetrapeptide.

\subsection{Sampling LJ$_{38}$ clusters} 

Calculations for the LJ$_{38}$ cluster are performed with the LAMMPS code \cite{plim95jcp}, with both ATLAS and META. We evaluate the FES at T=0.12 (expressed in reduced units), which is below the melting point of this system. This thermodynamic state point is characterized by the presence of two metastable crystalline structures -- one corresponding to a truncated octahedron with $fcc$ motifs, one associated with a defective icosahedron~\cite{wale03book}. The two structures can be discerned using the coordination number of each atom in the cluster.  The number of atoms with coordination number $c$ is expressed as
\begin{equation}
n_{c}=\sum_{i=1}^{N}e^{-\frac{\left(c_{i}-c\right)^{2}}{2\eta^{2}}},
\end{equation}
where the coordination $c_{i}$ for each atom is calculated as a function of the distance $d$ between them,
\begin{equation}
c_{i}=\sum_{j}\mathcal{S}\left(\left|\br_{i}-\br_{j}\right|\right),\qquad\mathcal{S}\left(d\right)=\begin{cases}
0 & d>r_{0}\\
1 & d<r_{1}\\
\left(y-1\right)^{2}\left(2y+1\right) & r_{1}<d<r_{0},\quad y=\frac{d-r_{1}}{r_{0}-r_{1}}
\end{cases}.
\end{equation}
For these simulations, we use the parameters $\eta=0.5$, $r_0=1.5$ and $r_1=1.25$ reduced units, respectively. 
We use a high dimensional description that involves 8 CVs that correspond to the numbers of atoms with coordination number ranging from 4 to 11. In all cases, after the generation of a satisfactory GMM, we use ATLAS with the two first principal components, i.e., 2D-PCA, that has proven to be very effective for the challenging FFH model potential. 
For META, we use a 2D bias based on $n_6$ and $n_8$, a pair of CVs which has been shown to be able to recover the FES in previous calculations \cite{wale03book,gibe+20jctc}. The complete list of parameters used in the ATLAS calculation is reported in the \SI.

\begin{figure}
    \centering
    \includegraphics[width=0.75\textwidth]{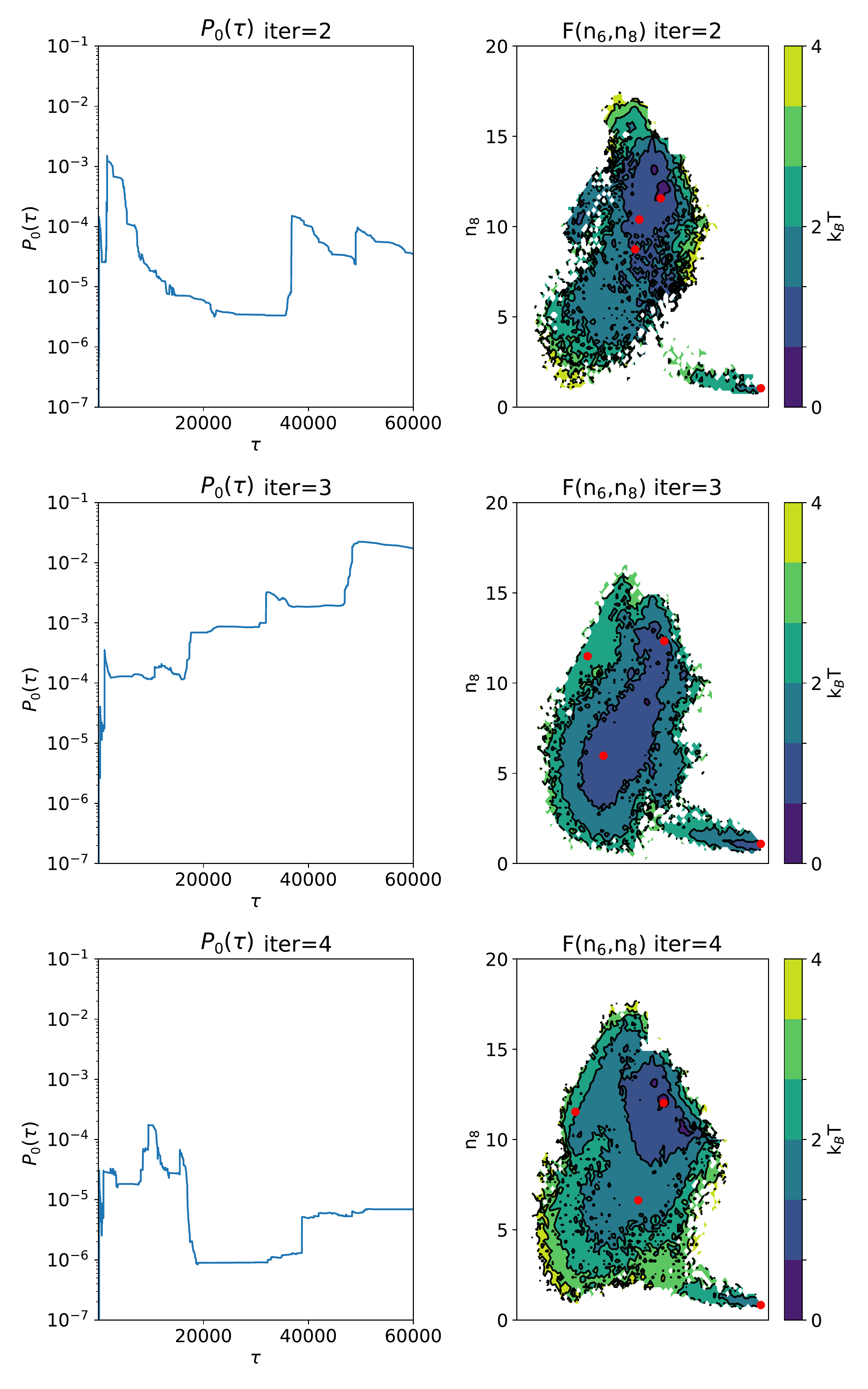}
    \caption{Behaviour of the first four biased ATLAS calculations performed during the self-learning iterative protocol for the LJ cluster. For each iteration, we illustrate the behaviour of $P_0(\tau)$ and the FES obtained by reweighing the calculation. The iterations start at two, since the first trajectory is not biased.}
    \label{fig:learning-states-LJ}
\end{figure}

The convergence of the iterative procedure to construct the GMM underlying ATLAS is illustrated in figure \ref{fig:learning-states-LJ}. The initial pool of trajectories samples primarily the crystalline structures, and so all the GMM centers are concentrated in that region. The ATLAS trajectory based on this GMM pushes the system outside these localized basins quickly, leading to jumps in $P_0$. Adding this trajectory to the pool gives excessive weight to the high-energy liquid-like configurations, and the ATLAS trajectory shows an even more rapidly increasing $P_0$. 
Having now accumulated samples in both the liquid-like and solid-like region, the GMM determines four clusters (discussed in more detail below), and the corresponding ATLAS trajectory maintains a value of $P_0$ below $10^{-5}$.

\begin{figure}
    \centering
    \includegraphics[width=\textwidth]{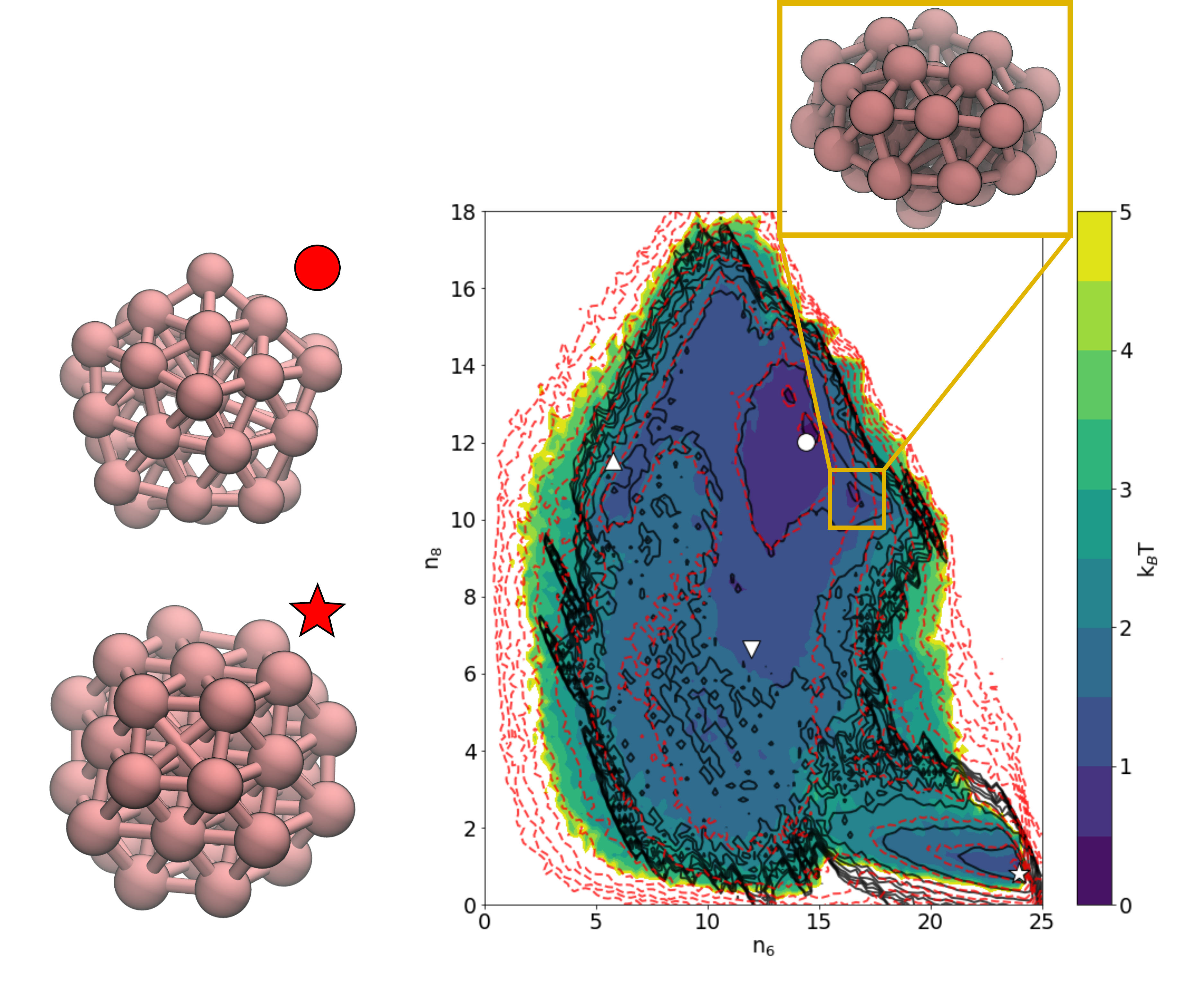}
    \caption{ATLAS and META FES obtained for the LJ-38. The low density structure, identified with $\color{red} \bullet$, and the high density one, identified as ${\color{red} \bigstar}$, are illustrated on the left of the FES. The ATLAS FES is shown with filled contour, the META FES with red dashed lines, and the PT reference with black thick lines. The minimum that META trajectory cannot recover is highlighted with an orange rectangle, and the corresponding structure is illustrated in an inset.} 
    \label{fig:LJ-FES-2var}
\end{figure}

We then run a longer 2D-PCA ATLAS trajectory based on this GMM to compute the FES. Figure~\ref{fig:LJ-FES-2var} shows the FES relative to $n_6$ and $n_8$, comparing the results obtained by applying ITRE to the ATLAS trajectory with those from the META and PT trajectories. Representative configurations for the clusters that correspond to the \emph{fcc} and icosahedral configurations are also depicted. 
The three FES are in good agreement with each other, but there are fine details that are not captured by the META trajectory. In particular, it does not identify a separate minimum, clearly present in both the PT and ATLAS trajectories, which corresponds to a more symmetric form of the icosahedral cluster.  This strucutre was also identified as a separate free energy minimum in a study using the dimensionality reduction algorithm sketch-map~\cite{ceri+13jctc}.  

\begin{figure}
    \centering
    \includegraphics[width=0.75\textwidth]{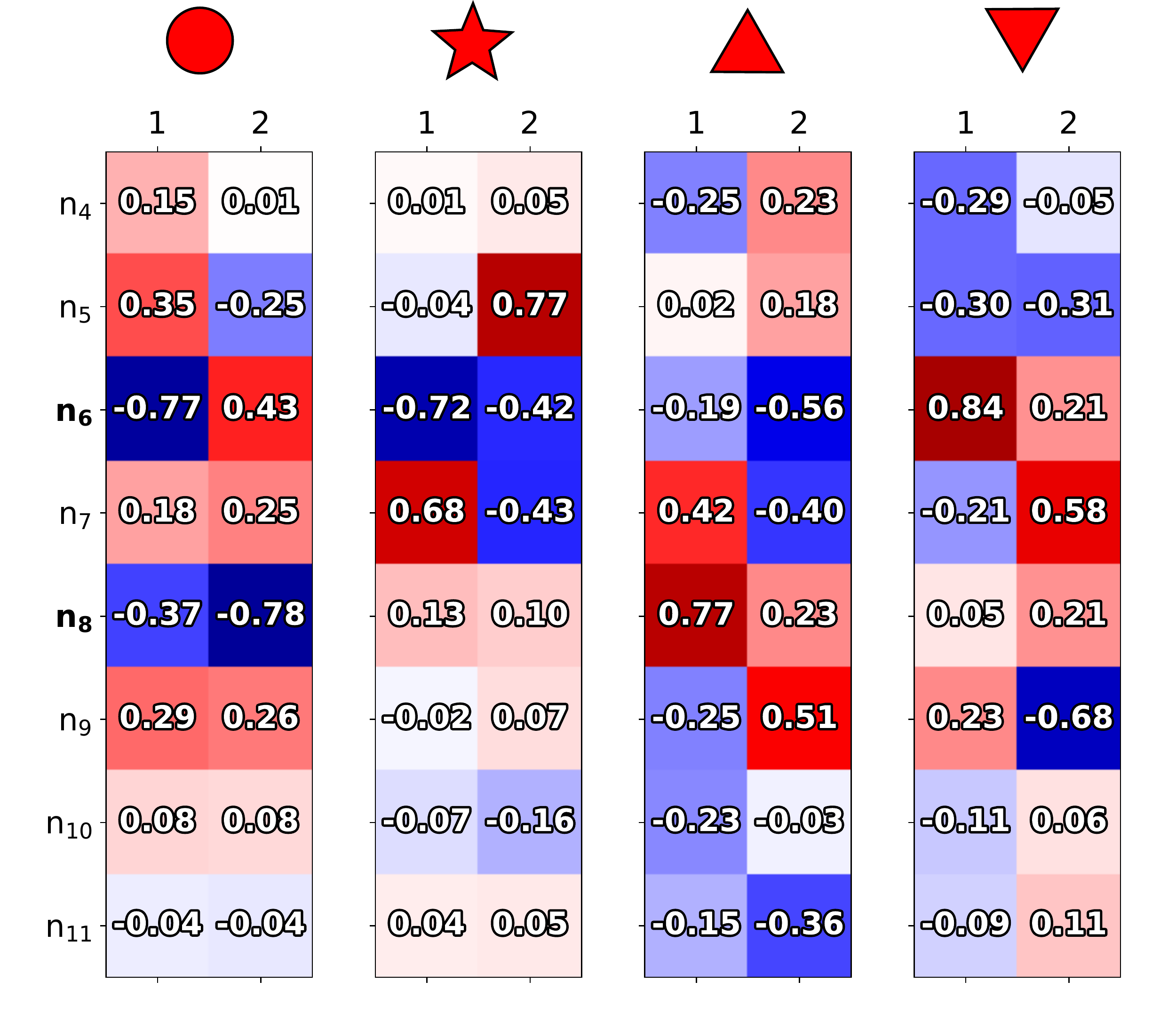}
    \caption{Color maps indicating the magnitude of the components of the top two eigenvectors $\bU_1$ and $\bU_2$ used in the ATLAS calculation to define the local CVs for each of the four clusters shown in Fig.~\ref{fig:LJ-FES-2var}. Each component in the eigenvectors indicates the importance of one of the coordination histogram bins $n_c$. The CVs used in the META calculations are highlighted in bold.}
    \label{fig:LJ-eigenvect}
\end{figure}

The accuracy of the ATLAS FES can be understood as a consequence of the better description of the local structure of each basin. Even though the free energy minima associated with the four clusters can be separated well using only  $n_6$ and $n_8$, the shape of the basins in the 8-dimensional CV space is not fully described by these two order parameters. 
This is seen clearly by analyzing the components of the eigenvectors $\bU_k$ associated with the local PCA that underlies the GMM, as shown in \ref{fig:LJ-eigenvect}. The histogram bins associated with $n_5$, $n_7$ and $n_9$ give equal or larger contributions to the top PCA components than $n_6$ and $n_8$. 
The adaptive topology described by the ATLAS bias, together with the efficiency of a low-dimensional local bias, enables a more efficient sampling of the phase space and a faster convergence in terms of both the qualitative features of the FES and of a quantitative measure given by the KL divergence (see \SI{}, Fig.~\ref{figSI:convergence-lj38}).

\subsection{Alanine oligopeptides}

To simulate the peptides, we use GROMACS-2018 and the Amber99sb forcefield\cite{abraham2015gromacs,pall2014tackling,pronk2013gromacs,lindorff2010improved,wang2004development}. Both alanine dipeptide and tetrapeptide are simulated at room temperature employing a stochastic velocity rescaling thermostat\cite{buss+07jcp}. For each system, we use the $\phi$ and $\psi$ dihedral angles of the backbone of the structure as CVs. This results in two angles for the dipeptide and six angles for the tetrapeptide. Since the CVs are periodic, rather than using Gaussian functions in the GMM, we employ mono-dimensional von-Mises distribution function as previously done by Gasparotto \textit{et al.} \cite{gasp+18jctc}
\begin{equation}
 G(\bmu, \bSigma | \bs) = \prod_i^D \frac{e^{\kappa_i cos(s_i-\mu_i)}}{2\pi I_0(\kappa_i)},
\end{equation}
In this expression $I_0(\kappa_i)$ is the zeroth-order modified Bessel function\cite{sra2012short}. Using this expression reduces the calculation of the total probability distribution function to the product of mono-dimensional Von-Mises distributions. Using the 2D-PCA framework, in this case means selecting the two largest $\kappa_i$ that characterize the distribution function. For simplicity, we use the two components with the higher $\kappa_i$ directly without normalizing for the eigenvalue, as this allows us to keep the same periodicity for all the $\bc$ variable (i.e. the same as the CVs).

\begin{figure}
    \centering
    \includegraphics[width=\textwidth]{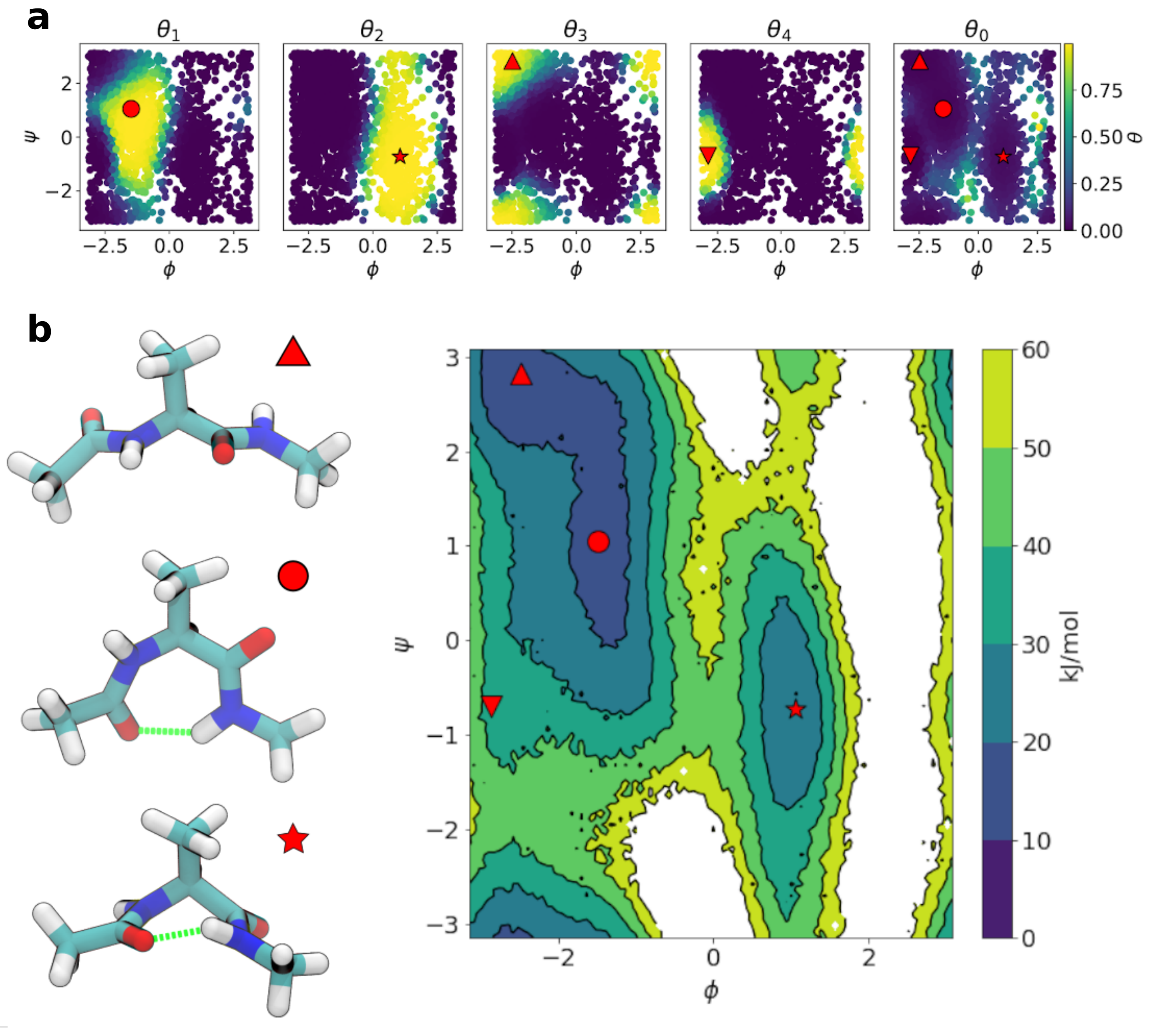}
    \caption{a) PMIs obtained using the Von-Mises distribution for the four clusters found for alanine-dipeptide, as well as the $\theta_0$ distribution. The PMIs can identify local motifs of the Ramachandran plot even across periodic boundary conditions, as it is clear for $\theta_2$, $\theta_3$, and $\theta_4$. b) FES obtained for the alanine dipeptide as a function of $\phi$ and $\psi$ angles. The three most stable structures are illustrated as $\color{red} \blacktriangle$ for the unfolded, as $\color{red} \bullet$ for the equatorial and $\color{red} \bigstar$ for the axial.}
    \label{fig:ala_dipept_pmis}
\end{figure}

We use alanine dipeptide as a proof of concept to illustrate the periodic version of ATLAS. We determine the GMM we use to converge the FES after four iterations of the same scheme we applied to LJ$_{38}$. This results in 4 clusters, three of which are associated with well-known minima in the FES of the dipeptide. The fourth is associated with a non-Gaussian feature on the landscape. 
It should be noted that even if the GMM identifies an additional cluster that does not correspond to a minimum, this does not corrupt the biasing scheme: in fact, the ATLAS basins provide a description of the CV space and do not necessarily need to be associated with a clear minimum.
As shown in the top panel of Figure~\ref{fig:ala_dipept_pmis}A, the PMIs based on  Von-Mises distributions identify the clusters correctly and account for the periodicity. The FES obtained by ITRE post-processing of the ATLAS sampling, shown in the bottom panel of the figure, corresponds to the well-known landscape for alanine dipeptide.

\begin{figure}
    \centering
    \includegraphics[width=0.75\textwidth]{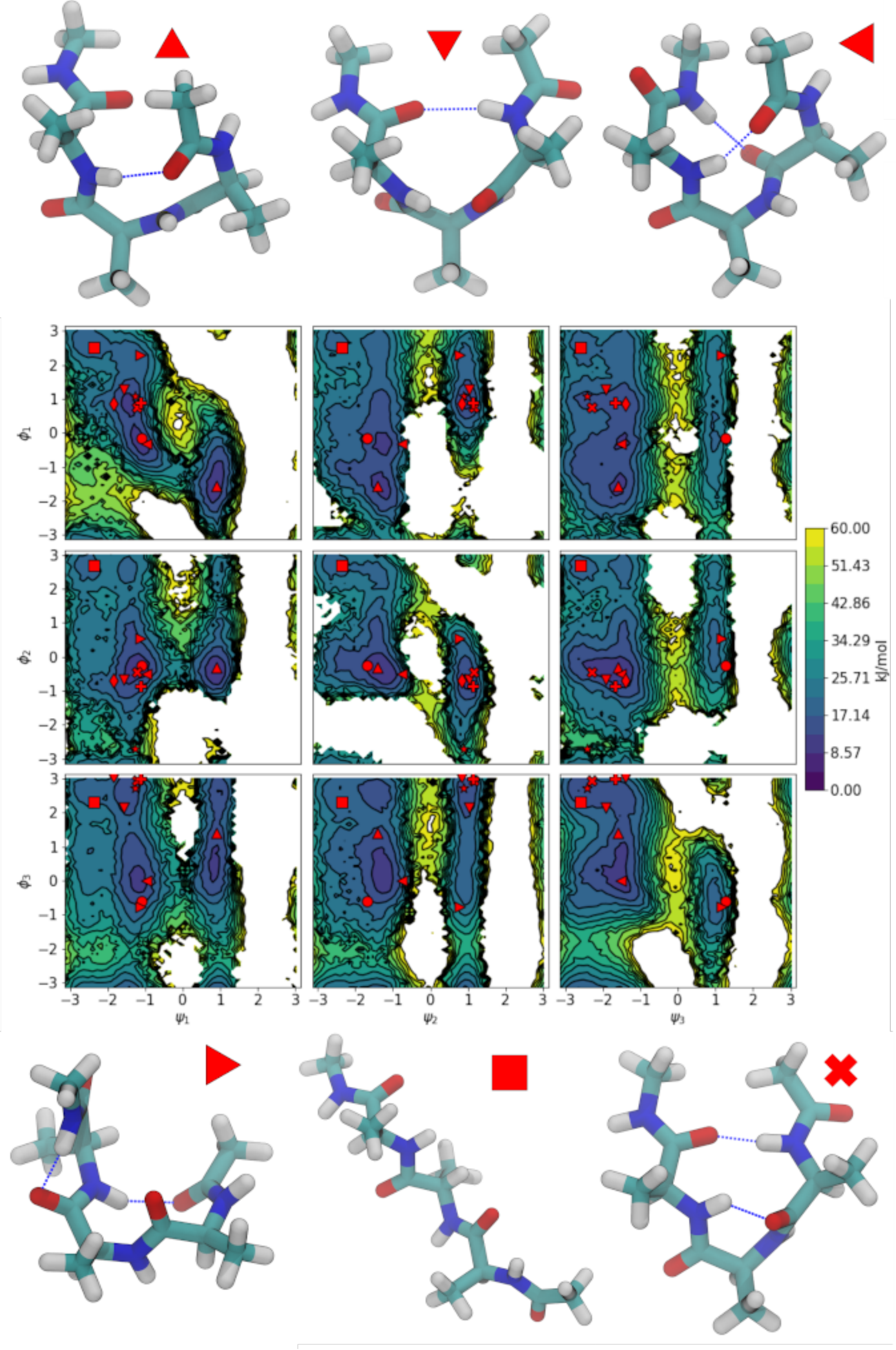}
    \caption{Projection of the FES of Alanine-tetrapeptide along nine different combinations of the backbone $\psi$-$\phi$ angles. Representative snapshots for the six most populated clusters in the GMM are also shown.}
    \label{fig:ala_tetrapept_fes}
\end{figure}

\begin{figure}
    \centering
    \includegraphics[width = \textwidth]{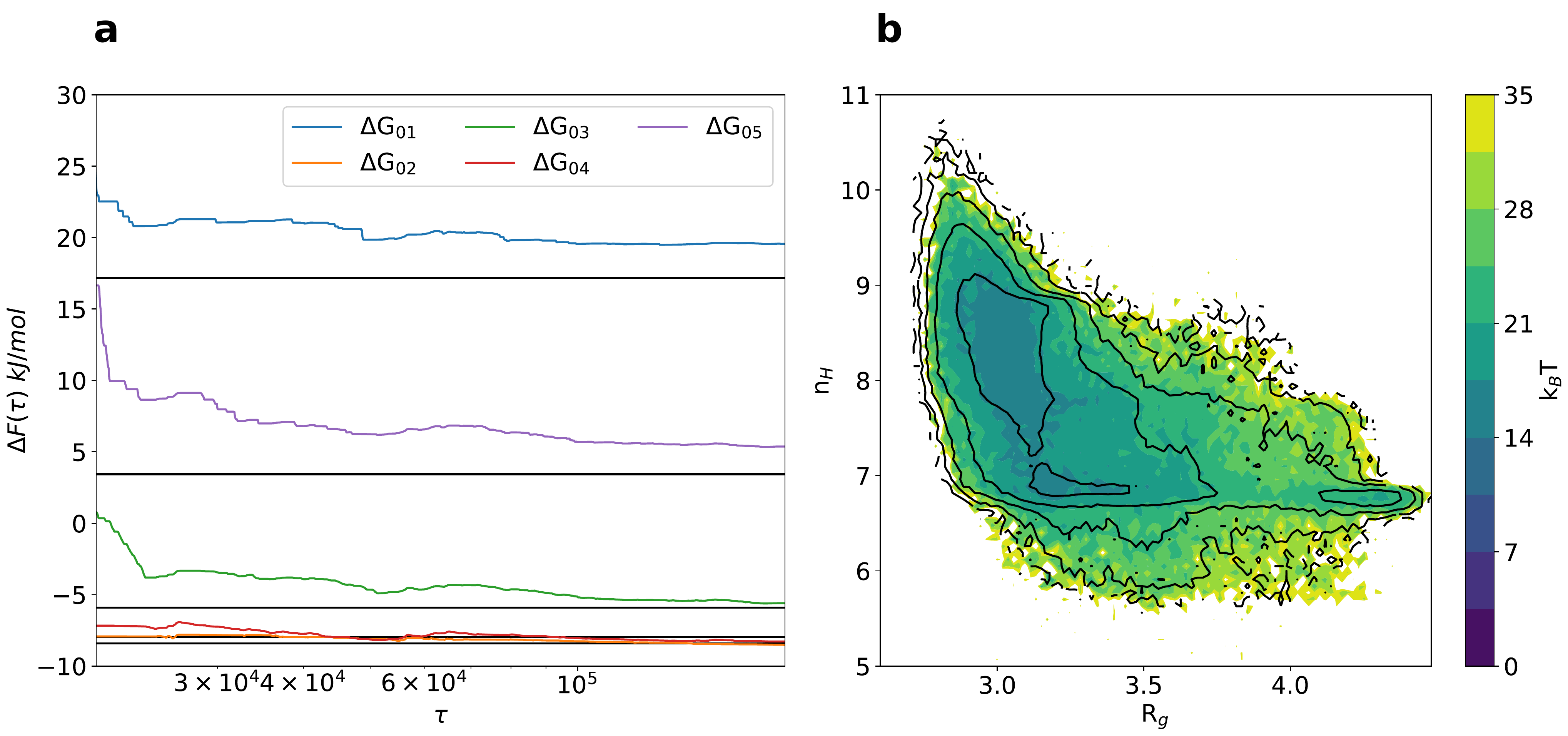}
    \caption{a) convergence of the free energy difference between the first basin and the six most populated basins identified for the Alanine tetrapeptide as a function of the deposition time $\tau=1 ps$, evaluated using equation \eqref{eq:delta_g}. b) converged FES as a function of the gyration radius R$_g$ and the number of hydrogen bonds of the backbone $n_H$, obtained by reweighting the ATLAS trajectory (filled colored contour) as well as the reference obtained with PT.}
    \label{fig:ala_tetrapept_fes_diff}
\end{figure}

While Ala$_2$ is a good demonstrative example, it cannot be considered a realistic, challenging example for the application of ATLAS. 
As a more complex, meaningful example we sample the six-dimensional dihedral space of alanine tetrapeptide. 
After five iterations of the self-learning algorithm, the GMM identifies a total of nine different clusters. Some of them correspond to well-defined minima, while others represent metastable states, with very low weight. We illustrate representative configurations from the six basins with the highest weights in figure \ref{fig:ala_tetrapept_fes}, together with nine FES obtained by reweighing nine different pairs of $\psi$-$\phi$ angles.
The free energy surface is smooth, and the system explores all the nine portions of phase space associated with the GMM very rapidly. ATLAS recovers the correct free energy differences between the different minima, as illustrated in Fig~\ref{fig:ala_tetrapept_fes_diff}a.

In order to quantitatively assess the performance of ATLAS in comparison with conventional well-tempered metadynamics calculations, we compare the 2D-PCA ATLAS calculations with two META simulations. 
In the first one we bias the same space as ATLAS, i.e. the six dihedral angles. 
As shown in the \SI{} (fig.~\ref{figSI:KL_mono_tetrapept} and \ref{figSI:KL_bi_tetrapept})  ATLAS converges to the reference PT FES more quickly, similar to what we observed for the 6D-FFH. 
The second metadynamics trajectory is representative of a typical metadynamics calculation, in which a low-dimensional bias is accumulated as a function of chemically-inspired CVs, namely the radius of gyration $R_g$ of the backbone and the number of hydrogen bonds $n_H$ between the carboxyl groups and the hydrogens of the amide groups.
Unsurprisingly, this simulation is even slower than the 6D-META trajectory in sampling the free-energy relative to the six dihedral angles (fig.~\ref{figSI:KL_mono_tetrapept} and \ref{figSI:KL_bi_tetrapept}).  It would seem that even for this simple system, two CVs are not enough to fully represent the configurational landscape. 
What is more, the ATLAS trajectory can be used to compute the FES as a function of $R_g$ and $n_H$. (fig.~\ref{fig:ala_tetrapept_fes_diff}b) shows that the estimate of this free energy from ATLAS converges faster than the estimate from the direct METAD sampling of those two CVs (see \ref{figSI:KL_gyr_tetrapept}). 
This example illustrates how the effective sampling of a high-dimensional configuration space is also beneficial when it comes to accelerating the convergence of the FES relative to conventional, easy-to-interpret CVs.

\section{Conclusions}
In this work, we have introduced the Adaptive Topography of Landscapes for Accelerated Sampling (ATLAS). We have demonstrated that this new sampling technique can enhance the sampling for a large number of degrees of freedom. 
The \textit{divide-et-impera} paradigm at the heart of ATLAS ensures that, at variance with conventional metadynamics whose cost scales exponentially with the number of CVs, the computational cost for ATLAS scales linearly with the number of distinct states that are accessible to the system. Biasing many degrees of freedom is thus feasible with ATLAS, making the choice of suitable CVs less critical. 

ATLAS's core idea is to partition a high-dimensional phase space into clusters, and to define  a suitable, low-dimensional local description of phase space in each of the identified regions. 
This description of phase space as a patchwork is appealing as it is in accord with our physical intuition of a free energy landscape composed of a series of basins connected by transition pathways. This physically intuitive representation for the bias makes it straightforward to interrogate the results of an ATLAS calculation. 
The local representations on which the bias acts can be used to understand the dynamics in the vicinity of the various stable states in the free energy landscape. Simultaneously, the PMIs provide a mechanism for understanding when the system has undergone a transition between two stable states. Therefore, the PMIs extracted from an ATLAS simulation could be used when fitting coarse-grained, models to describe the dynamics of the physical system. 

Although it is useful if the clusters in an ATLAS calculation correspond to basins in the energy landscape, this is not essential. We have shown how, as the calculation progresses and the system explores more of phase space, the locations of clusters can be recalculated, thereby refining the description of phase space that is used in the bias. As this refinement is achieved by applying an automated clustering step to the sampled trajectory, ATLAS automates the process of finding the best description of phase space and depositing the bias. ATLAS can thus be used to properly reconnoitre phase space for systems where one lacks physical intuition by extracting an intuitive coarse-grained representation for the high dimensional FES.

We also remark that the fundamental idea of a piecewise approximation of a complex free-energy landscape can be applied in more sophisticated ways than the one we discuss here. Assigning separate basins to transition states, using dynamics-aware definitions of the slow degrees of freedom within each basin, and applying different accelerated sampling schemes within each local map are all promising research directions, that could further facilitate the simulation of systems that involve several activated transitions that each require a different set of collective variables to achieve efficient statistical sampling.

\section{Acknowledgments}
FG, and MC were supported by the European Research Council under the European Union's Horizon 2020 research and innovation programme (Grant Agreement No. 677013-HBMAP), and by the NCCR MARVEL, funded by the Swiss National Science Foundation (SNSF).

\section{Supporting information}

Further diagnostic information on the simulations that are reported in this paper are available in the supporting information.  This information is available free of charge via the Internet at http://pubs.acs.org.

\providecommand{\latin}[1]{#1}
\makeatletter
\providecommand{\doi}
  {\begingroup\let\do\@makeother\dospecials
  \catcode`\{=1 \catcode`\}=2 \doi@aux}
\providecommand{\doi@aux}[1]{\endgroup\texttt{#1}}
\makeatother
\providecommand*\mcitethebibliography{\thebibliography}
\csname @ifundefined\endcsname{endmcitethebibliography}
  {\let\endmcitethebibliography\endthebibliography}{}

\end{document}